\documentclass[twocolumn,prd,nofootinbib]{revtex4}
\usepackage{multirow}
\usepackage{natbib}
\usepackage{xcolor}
\usepackage{bm}
\usepackage{amssymb}
\usepackage{amsmath}
\usepackage{braket}
\usepackage[para,online,flushleft]{threeparttable}
\usepackage{appendix}
\usepackage{hyperref}
\begin{document}

\title{Effects of $CP$-violating internucleon interactions in paramagnetic molecules}

\author{V. V. Flambaum$^{1,2}$}
\author{I. B. Samsonov$^1$}
\author{H. B. Tran Tan$^1$}
\affiliation{$^1$School of Physics, University of New South Wales,
Sydney 2052, Australia}
\affiliation{$^2$Helmholtz Institute Mainz, Johannes Gutenberg University, 55099 Mainz, Germany}

\begin{abstract}
    We demonstrate that electron electric dipole moment (eEDM) experiments with molecules in paramagnetic state are sensitive to $P,T$-violating nuclear forces and other $CP$-violating parameters in the hadronic sector. These experiments, in particular, measure the coupling constant $C_{SP}$ of the $CP$-odd contact semileptonic interaction. We establish relations between $C_{SP}$ and different $CP$-violating hadronic parameters including strength constants of the  $CP$-odd nuclear potentials, $CP$-odd pion-nucleon interactions, quark-chromo EDM and QCD vacuum angle. These relations allow us to find limits on various $CP$-odd hadronic parameters.
    %The obtained limits are comparable to the ones originating from nucleon EDM contribution to the atomic EDM.
\end{abstract}{}
%\date{\today}
\maketitle

\section{Introduction} 

%%%%%%%%%%%%%%%%%%%%%%%%%%%%%%%%%%%%%%%%%%%%%%%%%%%%%%%%%%%%%%%%%%%%%%%%%%%%%%%%%%%%%%

The existence of non-vanishing electric dipole moments (EDMs) of elementary particles was conjectured nearly seventy years ago \cite{Purcell1950,leeyang1957,LANDAU57a,LANDAU57b}, but they have not been observed so far. 
Their discovery would be a crucial step in the study of charge and parity ($CP$) violations.
On the one hand, the Standard Model of elementary particles predicts  non-vanishing values for the EDMs of the electron and nucleons, but these values are so minuscule that they are practically unobservable in current experiments. On the other hand, it is known that the $CP$-violation is needed to explain the apparent matter-antimatter asymmetry in the universe \cite{Sakharov91}. It is therefore an important challenge for experimental physics to measure the EDMs of elementary particles as well as EDMs of composite objects such as nucleons, nuclei and atoms. 

In the last decade or so, tremendous progress has been achieved in the experiments with paramagetic molecules which measure the electron EDM through specific energy level shifts \cite{ACMEII,Loh2013} (see also Refs.~\cite{yamanaka2017,chupp2019} for reviews). As was demonstrated in the recent works \cite{FPRS,FST2020}, these experiments are also sensitive to the $CP$-violating interactions in the hadronic sector, which originate, in particular, from the nucleon EDMs. The aim of the current work is to extend the results of the papers \cite{FPRS,FST2020} and to study the sensitivity of paramagnetic EDM experiments to $P,T$-violating nuclear forces.

It is a well-known fact that the $CP$-violating  effects are significantly enhanced in heavy atoms \cite{khriplovich1991parity}. In this paper, we consider a single-particle nuclear model in which the valence nucleon interacts with a heavy nuclear core through a nuclear potential $U({\bf r})$. The $P,T$-violating nuclear forces are taken into account by the phenomenological interaction Hamiltonian $H_{\rm odd} = \xi \boldsymbol{\sigma}\cdot\nabla U({\bf r})$, where $\boldsymbol{\sigma}/2$ is the nucleon spin operator and $\xi$ is the coupling constant of dimension of length. This coupling constant will be denoted by $\xi_p$ for proton and $\xi_n$ for neutron. The authors of Ref.~\cite{Schiffmoment1} demonstrated that the leading contributions to this interaction arise due to $\pi$ meson exchange between the valence nucleon and the nuclear core. In general, however, this interaction may arise due to the $CP$-violating $\pi NN$, $\eta NN$, three-pion and four-nucleon interactions. The last two interactions were considered, in particular, in Ref.~\cite{Bsaisou}.

In this paper, we will focus on the contributions to the atomic EDM arising due to the $P,T$-odd nuclear force with the Hamiltonian $H_{\rm odd}$ regardless of the underlying fundamental interaction. In principle, experimental limits on the phenomenological parameters $\xi_{p,n}$ may be converted into limits on the parameters of more fundamental hadronic interactions. We stress that the interaction $H_{\rm odd}$ considered in this work is an independent source of $CP$ violating effects, separate from the contributions due to nucleon EDMs considered in Refs.~\cite{FPRS,FST2020}.

The most stringent experimental constraint on the electron EDM was obtained by the ACME collaboration \cite{ACMEII} which measured specific energy level shifts in the $^{232}$ThO molecule. This experiment also placed a limit on the $CP$-odd electron-nucleon interaction coupling constant (90\% C.L.),
\begin{equation}
    \left|C_{SP}\right|_{\rm Th}
< 7.3 \times 10^{-10}\,.
\label{Csp-constr}
\end{equation}
In a heavy nucleus with $Z$ protons and $N=A-Z$ neutrons this coupling constant is a linear combination of independent electron couplings to protons ($C_{SP}^p$) and neutrons ($C_{SP}^n$), $C_{SP}= C_{SP}^p Z/A + C_{SP}^n N/A$. The latter two coupling constants correspond to the following $CP$-odd semileptonic operators
\begin{equation}\label{contact-interaction}
{\cal L} = \frac{iG_F}{\sqrt2}  C_{SP}^p\, \bar e  \gamma_5 e\,
\bar p p 
+ \frac{iG_F}{\sqrt2} C_{SP}^n\, \bar e  \gamma_5 e\,
\bar n n \,,
\end{equation}
where $G_F$ is the Fermi coupling constant, $e$, $p$ and $n$ are respectively the electron, proton and neutron fields. Note that the subscript `$SP$' denotes the nucleon-scalar and electron pseudoscalar two-fermion bilinears. 

Our goal is to establish the leading-order relation between the coupling constant $C_{SP}$ and the parameters of the $P,T$-odd internucleon interaction $\xi_p$ and $\xi_n$, $C_{SP} = C_{SP}(\xi_p,\xi_n)$. This will allow us to find limits on these couplings originating from the experimental constraint (\ref{Csp-constr}). Then, using known relations between the constants $\xi_p,\xi_n$ and more fundamental $CP$-violating hadronic parameters, we will establish leading-order relation between $C_{SP}$ and $CP$-odd pion-nucleon couplings $\bar g^{(0,1,2)}_{\pi NN}$, quark-chromo EDMs $\tilde d_{u,d}$ and QCD vacuum angle $\bar\theta$.

Note that this problem involves the third order perturbation theory in the nuclear part and second order in the electron-nucleus interaction. To get through these complications we have to make some approximations in the nuclear part of the problem where we perform the calculations analytically. While all these approximations are common and justifiable, we cannot pretend that the accuracy of our results is better than a factor of two. However, this accuracy is comparable to that in other calculations of the hadronic contributions to atomic EDM where the limits on the $CP$-violating parameters are often presented on the logarithmic scale. For instance, current limit on QCD vacuum angle reads $|\bar\theta| < 10^{-10}$, see, e.g., reviews \cite{yamanaka2017,chupp2019} and updated limits presented in Ref.~\cite{FD2020}. 
%The calculations in  the electron part of problem have been performed numerically and practically do not contribute to the errors of the final results.

The rest of the paper is organized as follows. In Sect.~\ref{Contributions to the atomic EDM from CP-odd nuclear forces}, we present an estimate for the atomic EDM arising from the $CP$-odd nuclear forces. In Sect.~\ref{Constraints}, we compare this contribution to the atomic EDM with that of the contact electron-nucleon interaction and the nucleon permanent EDMs and find relations between the constant $C_{SP}$ and $CP$-violating hadronic parameters. In Sect.~\ref{Conclusion} we give a summary of our results and provide some comments on assumptions and precision. Technical details of calculations of electronic and nuclear matrix elements are collected in appendices.

Throughout this paper we use natural units with $c=\hbar=1$.

%%%%%%%%%%%%%%%%%%%%%%%%%%%%%%%%%%%%%%%%%%%%%%%%%%%%%%%%%%%%%%%%%%%%%%%

\section{Contributions to the atomic EDM from \texorpdfstring{$P,T$}{}-odd nuclear forces}\label{Contributions to the atomic EDM from CP-odd nuclear forces}

In this section, we determine the contributions to the atomic EDM arising due to nuclear $P,T$-odd interactions. In Sect.~\ref{background} we start with a review of the $P,T$-perturbed nuclear wave functions, which were found in Ref.~\cite{Schiffmoment1}. These wave functions will be used in Sect.~\ref{PT-EDM} for the computation of the nuclear matrix elements contributing to the atomic EDM.

\subsection{Nuclear wave functions perturbed by \texorpdfstring{$P,T$}{}-odd nuclear interactions}
\label{background}
The nucleons in a nucleus can exhibit different $P,T$-odd interactions originating both from the Standard Model and beyond. Independently of their nature, in the non-relativistic limit, these short-range interactions may be taken into account by the following phenomenological single-particle Hamiltonian \cite{Schiffmoment1}
\begin{equation}
\label{H'}
H_{\rm odd} = \frac{G_F}{\sqrt2}\frac{\eta}{2m_p}\boldsymbol{\sigma} \cdot \nabla\rho({\bf r})\,,
\end{equation}
where $\boldsymbol{\sigma}/2$ is the spin of the valence non-relativistic nucleon, $\rho({\bf r})$ is the density of the nuclear core, $\eta$ is the effective coupling constant and $m_p$ is the proton mass. 

In a heavy nucleus, the nuclear core creates an effective nuclear potential $U({\bf r})$ in which the valence nucleon moves. In the short-range approximation, this potential is proportional to the density of the nuclear core, $\rho({\bf r})=U({\bf r})\rho(0)/U(0)$. Taking this into account, the Hamiltonian (\ref{H'}) may be represented as 
\begin{equation}\label{Hodd}
    H_{\rm odd}=\xi\boldsymbol{\sigma}\cdot \nabla U({\bf r})\,,
\end{equation}
where \cite{Schiffmoment1}
\begin{equation}
    \xi=\eta\frac{G_F}{2\sqrt2 m_p}\frac{\rho(0)}{U(0)}\approx -2\times 10^{-21} \eta\cdot {\rm cm}\,.
\end{equation}
The total potential for the valence nucleon is thus given by
\begin{equation}
    \tilde U = U + H_{\rm odd} = U + \xi\boldsymbol{\sigma}\cdot\nabla U \approx U({\bf r}+ \xi\boldsymbol{\sigma})\,.
\label{U+H}
\end{equation}

Let $|n'\rangle\equiv \psi_{n'}({\bf r})$ be unperturbed wave function of the valence nucleon labeled by some quantum numbers $n'$. This function is supposed to solve for the Schr\"odinger equation with the potential $U({\bf r})$. Equation (\ref{U+H}) suggests that the wave function purterubed by the $P,T$-odd interaction (\ref{H'}) may be represented as
\begin{equation}
    \tilde\psi_{n'}({\bf r}) = \psi_{n'}({\bf r}+ \xi\boldsymbol{\sigma})\approx (1+\xi\boldsymbol{\sigma}\cdot\nabla)\psi_{n'}({\bf r})\,,
\end{equation}
or, more generally, taking the sum over all nucleons
\begin{equation}
    |\tilde n'\rangle = \left(1+\sum_{i=1}^A\xi_i\boldsymbol{\sigma}_i\cdot\nabla_i\right) |n'\rangle\,.
    \label{CP-odd force}
\end{equation}
Note that the constants $\xi_i$ are different for proton ($\xi_i=\xi_p$) and neutron ($\xi_i=\xi_n$).

According to Eq.~(\ref{CP-odd force}), the matrix elements of an operator ${\cal O}$ may be written up to the first order in coupling constant $\xi$ as
\begin{equation}
\label{commutator}
    \langle \tilde m' | {\cal O} |\tilde n' \rangle
      = \langle m' | {\cal O} | n' \rangle 
      - \sum_{i=1}^A\xi_i \langle m' | [\boldsymbol{\sigma}_i\cdot \nabla_i ,{\cal O}] | n' \rangle\,.
\end{equation}

The $P,T$-perturbed wave functions (\ref{CP-odd force}) were found in Ref.\ \cite{Schiffmoment1}. In the next subsection we will use these wave functions to compute the nuclear matrix elements contributing to the atomic EDM.

\subsection{Electron-nucleon interaction Hamiltonian}
\label{PT-EDM}

%Our aim is to show that the $CP$-odd internucleon interaction \eqref{H'} yields the atomic EDM through electron-nucleon interaction. 
Let us consider a valence electron of charge $-e$ and position vector $\bf R$ interacting with a valence nucleon of charge $q$ which is located at the point $\bf r$. The interaction Hamiltonian considered in this paper is a combination of the electric ($q$) and magnetic ($\mu$) terms,
\begin{subequations}\label{mix_H}
\begin{align}
H_{\rm int}&=-H_q-H_\mu\,,\label{HInt}\\
H_q&= \frac{qe}{\left|{\bf R}-{\bf r}\right|}\,,\label{HE}\\
H_\mu&=\frac{e\,\boldsymbol{\mu}\cdot\left[\left({\bf R}-{\bf r}\right)\times\boldsymbol{\alpha}\right]}{\left|{\bf R}-{\bf r}\right|^3}\,,
\label{HB}
\end{align}
\end{subequations}
where 
\begin{equation}\label{magmoment}
    \boldsymbol{\mu}=\mu_0 (g^l {\bf l} + g^s {\bf s})
\end{equation} 
is the operator of nucleon's magnetic moment, $\mu_0$ is the nuclear magneton, $g^l$ and $g^s$ are the orbital and spin $g$-factors of the nucleon. Note that ${\bf s}=\frac12\boldsymbol{\sigma}$ is the nuclear spin operator while $\boldsymbol{\alpha}
=\left(
\begin{smallmatrix}
0 & \boldsymbol{\sigma} \\ \boldsymbol{\sigma} & 0 
\end{smallmatrix}
\right)$ are the Dirac matrices acting on electron's states. 

The interaction Hamiltonian (\ref{mix_H}) as a function of $\bf R$ and $\bf r$ may be expanded into a multipole series. In particular, the leading terms in the expansion of the electric interaction Hamiltonian (\ref{HE}) are 
\begin{equation}
\label{Hq}
\begin{aligned}
    H_q &= qe \left( \frac1R+ \frac{{\bf R}\cdot {\bf r}}{R^3} \right)\theta(R-r)\\
&+qe \left( \frac1{r}+ \frac{{\bf R}\cdot {\bf r}}{r^3} \right)\theta(r-R)
+\ldots\,,
\end{aligned}
\end{equation}
where the ellipsis stand for terms with higher multipolarity. In this expansion, the term $\theta(R-r)/R+\theta(r-R)/r$, after averaging with the nuclear charge density $\rho({\bf r})$, gives rise to the Coulomb interaction of the electron with the extended nucleus. This interaction is assumed to have already been taken into account by the unperturbed electronic wave functions. 

The term $({\bf R}\cdot {\bf r})[\theta(R-r)/R^{3}  +\theta(r-R)/r^{3} ]$ in Eq.~(\ref{Hq}) is the leading dipole one on which we will focus our attention. Evidently, at a distance from the nucleus, this term falls off as $1/R^2$. To find the effective electron-nucleon interaction Hamiltonian inside the nucleus, one needs to average this term over the normalized nuclear density, which, in the leading approximation, may be taken as constant inside the sphere of radius $R_0$ and vanishing outside, $\rho(r) = 3\theta(R_0-r)/R_0^3$. As a result, one obtains
\begin{equation}
\begin{aligned}
f(R)&\equiv  \int_0^{\infty} \rho(r)\left[ 
\frac1{R^2}\theta(R-r) + \frac1{r^2}\theta(r-R)
\right]r^2 dr\\
&=\frac1{R^2} \theta(R-R_0) + \frac{3R_0 - 2R }{R_0^3}\theta(R_0-R)\,.
\label{8}
\end{aligned}
\end{equation}
Using this prescription for the continuation of the $1/R^2$ function to small distances, one may extract the regularized dipole interaction from the operator (\ref{Hq}),
\begin{equation}
 \bar H_{q} \equiv qe (\hat{\bf R}\cdot {\bf r}) f(R)\,,
\label{Hq2}
\end{equation}
where $\hat{\bf R} \equiv {\bf R}/R$.

The magnetic interaction operator (\ref{HB}) also behaves as $1/R^2$ at large distances from the nucleus and may also be extended to the short-distance region inside the nucleus according to the prescription (\ref{8}),
\begin{equation}
    \bar H_\mu \equiv e\boldsymbol{\mu} \cdot (\hat{\bf R}\times \boldsymbol{\alpha})
f(R)\,.
\label{Hbarmu}
\end{equation}

Thus, the dipole part of the interaction (\ref{HInt}) which is regularized at short distances reads
\begin{equation}
    \bar H_{\rm int} = - \bar H_{q} - \bar H_\mu\,,
    \label{Hbarint}
\end{equation}
with $\bar H_{q}$ and $\bar H_\mu$ given by Eqs.~(\ref{Hq2}) and (\ref{Hbarmu}).

According to Eq.~(\ref{commutator}), to take into account the $P,T$-perturbed nuclear wave functions one has to consider the commutators of the operator $\boldsymbol{\sigma}\cdot\nabla_{\bf r}$ with the interaction Hamiltonians (\ref{Hq2}) and (\ref{Hbarmu}),
\begin{subequations}
\begin{align}
[\boldsymbol{\sigma}\cdot\nabla_{\bf r} ,\bar H_q] &= qe (\boldsymbol{\sigma}\cdot \hat{\bf R})
f(R)\,, \label{H1}\\ {}
[\boldsymbol{\sigma}\cdot\nabla_{\bf r} ,\bar H_\mu]&=
ie\mu_0 (g^s-g^l) f(R)(\hat{\bf R}\times \boldsymbol{\alpha})\cdot (\boldsymbol{\sigma}\times \nabla_{\bf r})\,.\label{H2}
\end{align}
\end{subequations}
In deriving Eq.~(\ref{H2}) we have applied the following commutator identities:
$[\sigma_i,\sigma_j] = 2i \varepsilon_{ijk}\sigma_k$ and $[\nabla_i,l_j ] = i \varepsilon_{ijk}\nabla_k$.

\subsection{Atomic EDM due to \texorpdfstring{$P,T$}{}-odd nuclear forces}

The unperturbed atomic states will be denoted by $|nn'\rangle= |n\rangle|n'\rangle$, where $|n\rangle$ and $|n'\rangle$ are electronic and nuclear states, respectively. In what follows, the nuclear quantum numbers will be distinguished from the electronic ones with the apostrophe. As in Sect.~\ref{Contributions to the atomic EDM from CP-odd nuclear forces}, the $P,T$-perturbed nuclear wave functions (\ref{CP-odd force}) are denoted as $|\tilde n'\rangle$.

The atomic EDM arising from the mixed interaction $H_{\rm int}$ may be calculated in perturbation theory. The first-order contribution to the atomic EDM vanishes for spinless nuclei which we consider in this paper,
\begin{equation}
      \sum_{n\ne0}
    \frac{\langle 0 |-e{\bf R}|n\rangle \langle n \tilde 0'|\bar H_{\rm int}|\tilde 0'0\rangle}{E_0-E_n} =0 \,.
\end{equation}
Indeed, it may be shown that $\langle \tilde 0' | \bar H_{\rm int} | \tilde 0'\rangle
\propto \langle 0' | {\bf s} | 0' \rangle$=0, where ${\bf s}$ is the nuclear spin operator.

The leading non-vanishing contributions to the atomic EDM thus arise in the second-order perturbation theory,
\begin{widetext}
\begin{align}
{\bf d}&= 2\sum_{m\ne0,n\tilde n'\ne 0\tilde0'}\frac{\bra{0}-e\mathbf{R}\ket{m}\bra{m\tilde0'}\bar H_{\rm int}\ket{\tilde n'n}\bra{n\tilde n'} \bar H_{\rm int}\ket{\tilde0'0}}{\left(E_{m}-E_{0}\right)[ E_{n}-E_0+{\rm sgn}(E_n)( E_{\tilde n'}-E_{\tilde 0'})]}\label{2.17} \\ 
&+\sum_{m\ne0,n\tilde n'\ne 0\tilde0'}
\frac{\langle 0\tilde0'| \bar H_{\rm int} | \tilde n'n\rangle
\langle n| -e{\bf R}|m\rangle
\langle m\tilde n'| \bar H_{\rm int}|\tilde 0'0\rangle}{
[E_n-E_0+{\rm sgn}(E_n)(E_{\tilde n'}-E_{\tilde 0'})][E_m-E_0+{\rm sgn}(E_m)(E_{\tilde n'}-E_{\tilde 0'})]}\,,
\nonumber
\end{align}
\end{widetext}
where the ${\rm sgn}(E_n)$ in the denominators is needed to correctly account for the negative energy electronic states. Indeed, the negative energy electronic states contribute with opposite sign of the nuclear energy because they may be viewed as blocking contributions which prevent the valence electron from directly transitioning into such states which are supposed to be completely occupied in the Dirac sea picture.

The term in the second line in Eq.~(\ref{2.17}) may be neglected in comparison with the other one because it is suppressed by higher power of nuclear energy in the denominator. Moreover, we assume that the leading contributions to the atomic EDM arise from the matrix elements with $|0\rangle = |s_{1/2}\rangle$ and $|m\rangle = |p_{1/2}\rangle$ electronic states because these wave functions are significantly enhanced in the vicinity of a heavy nucleus. Taking this into account, Eq.~(\ref{2.17}) may be cast in the form
\begin{equation}
\label{atomicEDM}
    {\bf d}\approx 2 \frac{\langle s_{1/2}|e {\bf R}|p_{1/2}\rangle \langle p_{1/2} | H_{\rm eff}| s_{1/2}\rangle }{E_{p_{1/2}}-E_{s_{1/2}}}\,,
\end{equation}
where we have introduced the effective interaction Hamiltonian
\begin{equation}
\label{Heff}
    \hspace{-0.5mm}H_{\rm eff} \equiv 
    \sum_{n\tilde n'\ne 0\tilde0'}
    \frac{|m\rangle\langle m \tilde 0'| \bar H_{\rm int}| \tilde n' n\rangle \langle n\tilde n'|\bar H_{\rm int}|\tilde 0' 0\rangle\langle 0|}{
    \Delta E_n +{\rm sgn}(E_n) \Delta E_{\tilde n'}}\,.
\end{equation}
Here $\Delta E_n = E_n - E_0$ and $\Delta E_{\tilde n'} = E_{\tilde n'} - E_{\tilde 0'}$. Our goal is to calculate the matrix element $\langle p_{1/2} | H_{\rm eff}| s_{1/2}\rangle$ which is responsible for the leading-order contributions to the atomic EDM.

Using the identity (\ref{commutator}), one may single out the part of the matrix element of the effective Hamiltonian (\ref{Heff}) which is linear in the $P,T$-odd nuclear interaction coupling $\xi$
\begin{widetext}
\begin{equation}\label{22}
  \langle p_{1/2} | H_{\rm eff}| s_{1/2}\rangle = -\xi
    \sum_{nn'\ne 00'}
    \frac{\langle p_{1/2}0'| [\boldsymbol{\sigma}\cdot\nabla, \bar H_{\rm int}]|  n' n\rangle \langle n n'|\bar H_{\rm int}| 0' s_{1/2}\rangle}{
    \Delta E_n + {\rm sgn}(E_n)\Delta E_{ n'}} + (s_{1/2}\leftrightarrow p_{1/2})\,.
%   \nonumber\\&&
% -\xi
%    \sum_{nn'\ne 0\tilde0'}
%    \frac{\langle p_{1/2}  0'|  \bar H_{\rm int}|  n' n\rangle \langle n n'|[\boldsymbol{\sigma}\cdot\nabla,\bar H_{\rm int}]| 0' s_{1/2}\rangle}{
%    \Delta E_n + {\rm sgn}(E_n)\Delta E_{ n'}}    \,.
\end{equation}
\end{widetext}
Here, for brevity, we use the generic symbol $\xi$ to uniformly denote $\xi_{p}$ and $\xi_n$. The specific proton and neutron contributions will be displayed explicitly in the final result. 

Substituting Eq.~(\ref{Hbarint}) into Eq.~(\ref{22}), we express this matrix element in terms of the operators (\ref{Hq2}) and (\ref{Hbarmu}),
\begin{equation}
  \langle p_{1/2} | H_{\rm eff}| s_{1/2}\rangle = 
 \sum_{nn'\ne 00'}
    \frac{\xi\left({\cal M}^1_{nn'} + {\cal M}^2_{nn'}\right)}{
    \Delta E_n + {\rm sgn}(E_n)\Delta E_{ n'}}\,,  
\label{eq2.21}
\end{equation}
where 
\begin{subequations}
\begin{align}
 {\cal M}^1_{nn'} &\equiv \langle p_{1/2}0'| [\boldsymbol{\sigma}\cdot\nabla, \bar H_q]|  n' n\rangle \langle n n'|\bar H_\mu| 0' s_{1/2}\rangle\,,\label{M1}\\
{\cal M}^2_{nn'} &\equiv\langle p_{1/2}0'| [\boldsymbol{\sigma}\cdot\nabla, \bar H_{\mu}]|  n' n\rangle \langle n n'|\bar H_q| 0' s_{1/2}\rangle\,. \label{M2}   
\end{align}
\end{subequations}
These matrix elements will be calculated in the next subsection.

%%%%%%%%%%%%%%%%%%%%%%%%%%%%%%%%%%%%%%%%%%%%%%%%

\subsection{Calculation of matrix elements}

Consider the matrix elements in Eq.~(\ref{M1}). Using the identities (\ref{Hbarmu}) and (\ref{H1}), one may separate its electronic and nuclear components as
\begin{equation}
\begin{aligned}
     {\cal M}^1_{nn'} &= qe^2 \langle 0'|\boldsymbol{\sigma} |n'\rangle\langle n' |\boldsymbol{\mu} | 0'\rangle\\
    &\times\langle p_{1/2} | f(R) \hat {\bf R} | n\rangle \langle n |f(R) \hat {\bf R}\times \boldsymbol{\alpha} |s_{1/2}\rangle\,.
\label{2.23}   
\end{aligned}
\end{equation}

Note that in the product of the nuclear matrix elements we may single out the scalar term which gives dominant contribution in spinless nuclei,
\begin{equation}
    \langle 0'|\sigma_i |n'\rangle \langle n' |\mu_j | 0'\rangle =  \frac13 \delta_{ij}\langle 0'|\boldsymbol{\sigma} |n'\rangle \langle n' |\boldsymbol{\mu} | 0'\rangle + \dots\,,
    \label{2.24}
\end{equation}
where the ellipsis stands for the tensor terms which we omit in further calculations. With the use of the definition (\ref{magmoment}), the expression $\langle 0'|\boldsymbol{\sigma} |n'\rangle \langle n' |\boldsymbol{\mu} | 0'\rangle$ reduces to the nuclear spin-flip matrix element
\begin{equation}
    \langle 0'|\boldsymbol{\sigma} |n'\rangle \langle n' |\boldsymbol{\mu} | 0'\rangle
    =2\mu_0 (g^s - \epsilon g^l )
    | \langle 0'|{\bf s} |n'\rangle|^2\,,
    \label{2.25}
\end{equation}
where 
\begin{equation}
    \epsilon = \left\{ 
      \begin{array}{l}
        1 \mbox{ for spherical nuclei} \\
        0 \mbox{ for deformed nuclei.}
      \end{array}
    \right.
\end{equation}
In Eq.~(\ref{2.25}), we have taken into account the fact that for spherical nuclei, the states are usually represented in the $lj$-basis in which $\langle 0'|{\bf s} |n'\rangle \langle n' |{\bf j} | 0'\rangle =0$ while the states of deformed nuclei are usually represented in the $ls$-basis with $\langle 0'|{\bf s} |n'\rangle \langle n' |{\bf l} | 0'\rangle =0$.

With the use of Eqs.~(\ref{2.24}) and (\ref{2.25}) the matrix element (\ref{2.23}) may now be cast in the form 
\begin{equation}
\begin{aligned}
{\cal M}^1_{nn'} &= \frac23 \mu_0 (g^s - \epsilon g^l ) qe^2 |\langle 0'|{\bf s} |n'\rangle|^2\\
&\times \langle p_{1/2} | f(R) \hat {\bf R} | n\rangle \langle n |f(R) \hat {\bf R}\times \boldsymbol{\alpha} |s_{1/2}\rangle\,.
\label{M1res}
\end{aligned}
\end{equation}

Similarly, one may write the expression~(\ref{M2}) for ${\cal M}^2_{nn'}$ as
\begin{equation}
 \begin{aligned}
    {\cal M}^2_{nn'} &= \frac i3e^2 q \mu_0 (g^s-g^l) \langle 0'|\boldsymbol{\sigma}\times \nabla_{\bf r} |n'\rangle  \langle n' |{\bf r} | 0'\rangle\\
    &\times\langle p_{1/2} | f(R) \hat {\bf R}\times\boldsymbol{\alpha}| n \rangle  \langle n| f(R) \hat{\bf R} |s_{1/2} \rangle\,.
\label{2.28}
\end{aligned}   
\end{equation}

Here $\langle n' |{\bf r} | 0'\rangle$ is the E1 nuclear transition matrix element which may be considered within the giant dipole resonance model. Effectively, this means that 
%Victor
the sum over $n'$ is dominated by the matrix elements $\langle n' |{\bf r} | 0'\rangle$ which constitute the giant electric dipole resonance with the excitation energy $\Delta\bar E$.
%for different $n'$ allmatrix elements $\langle n' |{\bf r} | 0'\rangle$ have roughly the same value corresponding to the energy of giant dipole resonance $\Delta\bar E$. Upon this assumption,
Then having fixed the nuclear energy in the denominator of Eq.~(\ref{eq2.21}),  one might use the completeness relation for the nuclear states, $|n'\rangle \langle n'| =1$, to reduce the nuclear matrix elements in Eq.~(\ref{2.28}) to the expectation value of the ${\bf l}\cdot {\bf s}$ operator,
\begin{equation}
    \langle 0'|\boldsymbol{\sigma}\times \nabla_{\bf r} |n'\rangle \langle n' |{\bf r} | 0'\rangle \approx -2i \langle 0' | {\bf l}\cdot{\bf s}  |0'\rangle \equiv -2i \langle {\bf l}\cdot{\bf s}  \rangle\,.
\end{equation}
With this expression for the nuclear matrix element, Eq.~(\ref{2.28}) simplifies to
\begin{equation}
\begin{aligned}
        {\cal M}^2_{nn'} &= \frac23e^2 q \mu_0 (g^s - g^l)  \langle 0'| {\bf l}\cdot{\bf s} |0' \rangle\\
&\times\langle p_{1/2} | f(R) \hat {\bf R}\times\boldsymbol{\alpha}| n \rangle \langle n| f(R) \hat{\bf R} |s_{1/2} \rangle
\,.
 \label{M2res}
\end{aligned}
\end{equation}

Substituting Eqs.~(\ref{M1res}) and (\ref{M2res}) into Eq.~(\ref{eq2.21}), one may represent the matrix element of the effective operator (\ref{Heff}) in the compact form
\begin{widetext}
\begin{equation}
    \langle p_{1/2} | H_{\rm eff}| s_{1/2}\rangle = 2\xi q \mu_0 \left[ \sum_{n'}(g^s-\epsilon g^l) M(E_{n'}) |\langle 0'| {\bf s}| n'\rangle|^2 + (g^s-g^l) M(\Delta\bar E) \langle 0' |{\bf l}\cdot{\bf s} |0'\rangle \right]\,,
    \label{2.33}
\end{equation}
\end{widetext}
where
\begin{align}
  M(E)& \equiv \frac{e^2}{3}\sum_{n}
    \frac{\langle p_{1/2}| f(R)\hat {\bf R}|n\rangle \langle n |f(R)\hat {\bf R}\times \boldsymbol{\alpha}|s_{1/2}\rangle}{\Delta E_n + {\rm sgn}(E_n) E}\nonumber\\
    &+(s_{1/2} \leftrightarrow p_{1/2})\,.
    \label{M(E)}  
\end{align}

Note that the sum in Eq.~(\ref{2.33}) contains only single-particle nucleon excitations. It is instructive to separate proton ($p$) and neutron ($n$) contributions with nuclear excitation energies denoted by $\Delta E_p$ and $\Delta E_n$ as 
\begin{equation}
\begin{aligned}
    &\langle p_{1/2} | H_{\rm eff}| s_{1/2}\rangle \\
    &= 2  \mu_0 \sum_{i=p,n}\xi_iq_i\left[
     (g^s_i - \epsilon g^l_i) M_i + (g^s_i-g^l_i) \bar{M}
    \langle {\bf l}\cdot{\bf s} \rangle_i \right]\,,
    \label{<H>}
    \end{aligned}
\end{equation}
where 
\begin{align}
 M_p &\equiv \sum_{\Delta E_p} |\langle 0'| {\bf s}| n'\rangle_p|^2 M(\Delta E_p)\,,\\
 M_n &\equiv \sum_{\Delta E_n} |\langle 0'| {\bf s}| n'\rangle_n|^2 M(\Delta E_n)\,,\\
 \bar{M}&\equiv M(\Delta\bar{E})\,.
\end{align}
We recall that the nucleon $g$-factors are $g^l_p = 1 $, $g^s_p = 5.586$ for proton and $g^l_n = 0$, $g^s_n = -3.826$ for neutron. The effective nucleon charge is modified by the recoil effect: $q=q_p\equiv e N/A$ for proton and $q=q_n \equiv - e Z/A$ for neutron.

%%%%%%%%%%%%%%%%%%%%%%%%%%%%%%%%%%%%%%%%%%

\subsection{Matrix elements of the effective Hamiltonian for some heavy atoms}

In this section, we present the results of numerical calculation of the matrix element (\ref{<H>}) for different heavy atoms of experimental interest including $^{138}$Ba, $^{206}$Pb, $^{208}$Pb, $^{172}$Yb, $^{174}$Yb, $^{176}$Yb, $^{178}$Hf, $^{180}$Hf, $^{226}$Ra, $^{232}$Th. These atoms, as parts of various paramagnetic molecules, have been considered or are proposed for consideration in recent and future eEDM experiments. 

The expression (\ref{<H>}) depends on different nuclear matrix elements and corresponding energies of nuclear transitions. In particular, $\langle 0'| {\bf s}| n'\rangle_p$ and $\langle 0'| {\bf s}| n'\rangle_n$ are matrix elements for nuclear spin-flip proton and neutron transitions with energies $\Delta E_p$ and $\Delta E_n$, respectively. These matrix elements and energies may be estimated within the Nilsson nuclear model \cite{BM} which takes into account single-particle excitations only. This model allows one to estimate also the expectation value of the ${\bf l}\cdot{\bf s}$ operator for proton $\langle{\bf l}\cdot{\bf s}\rangle_p$ and neutron $\langle{\bf l}\cdot{\bf s}\rangle_n$ states. The details of calculation of these matrix elements and the corresponding energies are given in Appendix~\ref{NuclearSection}, see Tables \ref{sphericalNuc} and \ref{deformNuc}. These tables contain also the energies of giant dipole resonance $\Delta\bar E$ which enter in the last term in Eq.~(\ref{<H>}).

The sum over the intermediate electronic states in Eq.~(\ref{<H>}) is taken into account with the function (\ref{M(E)}) which should be evaluated for each nuclear energy. This function involves electronic bound states $|s_{1/2}\rangle$ and $|p_{1/2}\rangle$, as well as intermediate excited electronic states $|n\rangle$. For simplicity, the intermediate electronic states are restricted to the continuum because the states in the discrete spectrum may be shown to give negligible contributions (see, e.g., \cite{Plunien91,Plunien95}). 

Note that the operators in the matrix elements in Eq.~(\ref{<H>}) are short-range because the function (\ref{8}) falls off as $1/R^2$ outside the nucleus. Therefore, these matrix elements receive their main contributions from the region $0<R\ll a_B/Z^{1/3}$, where $a_B$ is the Bohr radius. In this region, the inter-electron interaction and screening are negligible as compared with the electron-nucleus Coulomb interaction. Therefore, the states $|s_{1/2}\rangle$ and $|p_{1/2}\rangle$ may be described by the unscreened Dirac-Coulomb wave functions which are appropriately regularized inside the nucleus; see Appendix \ref{DiracCoilombF} for further details.% Note that in these wave functions we do not fix the normalization coefficients $c_{s_{1/2}}$ and $c_{p_{1/2}}$. Therefore, these coefficients will be present explicitly in the resulting expression for the matrix element (\ref{<H>}).

The intermediate electronic states $|n\rangle$ in Eq.~(\ref{M(E)}) are given by the Dirac-Coulomb wave functions in the continuous spectrum (see Appendix~\ref{AppB2}). Using these wave functions, we calculate numerically the radial integrals in the matrix elements (\ref{M(E)}) for each particular nuclear energy (see Appendix \ref{AppB3}). The results of these calculations are collected in Table~\ref{Electronic}. 

Using the values of the nuclear matrix elements from Tables \ref{sphericalNuc} and \ref{deformNuc}, and the values of the electronic matrix elements from Table \ref{Electronic}, we find the matrix element (\ref{<H>}) for various atoms,
\begin{equation}
    \langle p_{1/2} | H_{\rm eff} | s_{1/2} \rangle = 2 c_{s_{1/2}} c_{p_{1/2}}  \frac{e\mu_0}{a_{B}} 
    (\tilde\lambda_p \xi_p  + \tilde\lambda_n \xi_n)\,,
    \label{H-result}
\end{equation}
where $c_{s_{1/2}}$ and $c_{p_{1/2}}$ are the normalization coefficients of the wave functions (\ref{B2}) and
\begin{align}\label{lambda-coefs}
\tilde\lambda_p &= \frac{A-Z}{A} [(g^s_p - \epsilon g^l_p)M_p + (g^s_p-g^l_p) \bar M \langle {\bf l}\cdot{
\bf s}\rangle_p]\,,\\
\tilde\lambda_n &= -\frac{Z}{A} [(g^s_n - \epsilon g^l_n)M_n + (g^s_n-g^l_n) \bar M \langle {\bf l}\cdot{\bf s}\rangle_n]\,.
\end{align}
Numerical values of these coefficients are given in Table \ref{tab:lambda-pn}. Equation (\ref{H-result}) represents one of the main results on this paper as it specifies the leading-order dependence of the atomic EDM (\ref{atomicEDM}) on the $P,T$-odd coupling constants $\xi_p$ and $\xi_n$.
\begin{widetext}
\begin{center}
    \begin{table}[tbh]
    \centering
\begin{tabular}{|c|c|c|c|c|c|c|c|c|c|c|}
\hline
                  & ${}^{138}{\rm Ba}$ & ${}^{206}{\rm Pb}$ & ${}^{208}{\rm Pb}$ & ${}^{172}{\rm Yb}$ & ${}^{174}{\rm Yb}$ & ${}^{176}{\rm Yb}$ & ${}^{178}{\rm Hf}$ & ${}^{180}{\rm Hf}$ & ${}^{226}{\rm Ra}$ & ${}^{232}{\rm Th}$ \\
\hline
$\tilde\lambda_p/100$     & 0.65  & 7.0  & 7.1 & 3.6 & 3.6 & 3.6 & 5.4 & 5.4 & 11 & 15 \\
\hline
$\tilde\lambda_n/100$     & 0.68  & 5.8  & 4.8 & 2.4 & 2.9 & 2.3 & 2.6 & 3.2 & 6.7 & 11\\
\hline
\end{tabular}
    \caption{Results of numerical calculations of coefficients $\tilde\lambda_p$ and $\tilde\lambda_n$ which specify the leading-order dependence of the matrix element (\ref{H-result}) on $P,T$-odd coupling constants $\xi_p$ and $\xi_n$.}
    \label{tab:lambda-pn}
\end{table}
\end{center}
\end{widetext}

%%%%%%%%%%%%%%%%%%%%%%%%%%%%%%%%%%%%%%%%%%%%%%%%%%%%%%%%%%%%%%%%%%%%%%%

\section{Comparison with the contact \texorpdfstring{$CP$}{}-odd electron-nucleon interaction}
%Constraints on \texorpdfstring{$CP$}{}-odd hadronic parameters}
\label{Constraints}

In this section, we will compare the matrix element (\ref{H-result}) with that of the contact interaction (\ref{contact-interaction}). This will allow us to determine the dependence of the coupling constant $C_{SP}$ on the $P,T$-odd nuclear force coupling constants $\xi_p$ and $\xi_n$. Then, employing the experimental constraint (\ref{Csp-constr}) we will determine the limits on $\xi_{p}$ and $\xi_n$ originating from the EDM experiments with paramagnetic atoms and molecules. %We will also present the constraints on other hadronic parameters, which follow from the obtained ones.

%We start this section with a brief review of the contact interaction matrix element calculated in \cite{Our}.

\subsection{Limits on \texorpdfstring{$P,T$}{}-odd nuclear interaction couplings}

In an atom, the contact interaction (\ref{contact-interaction}) yields the following interaction Hamiltonian between a valence electron and a nucleus \cite{NewBounds1985}
\begin{equation}
H_{\rm cont}=\frac{iG_F}{\sqrt{2}}AC_{SP}\gamma_0\gamma_5 \rho({\bf R})\,,
\label{Hcont}
\end{equation}
where $\gamma_0$ and $\gamma_5$ are the Dirac matrices and $\rho({\bf R})$ is the normalized nuclear charge density. The matrix element of this operator with the $s_{1/2}$ and $p_{1/2}$ states was calculated in Ref.~\cite{FST2020},
\begin{equation}   
\begin{aligned}
    \bra{p_{1/2}}H_{\rm cont}\ket{s_{1/2}}& = -c_{s_{1/2}}c_{p_{1/2}}\frac{G_F C_{SP}}{10\sqrt2\pi}\\
   & \times\frac{1+4\gamma}{\Gamma(2\gamma+1)^2}\frac{AZ\alpha}{R_0^2} 
\left(\frac{2ZR_0}{a_B}\right)^{2\gamma},
\label{matrix-contact}
\end{aligned}
\end{equation}
where $\gamma = \sqrt{1-Z^2\alpha^2}$ is the relativistic factor.

Let us now compare the matrix elements (\ref{H-result}) and (\ref{matrix-contact}). Setting $\bra{p_{1/2}}H_{\rm cont}\ket{s_{1/2}} = \bra{p_{1/2}}H_{\rm eff}\ket{s_{1/2}}$ allows us to find the leading-order dependence of the contact interaction constant $C_{SP}$ on the $P,T$-odd nuclear interaction couplings $\xi_p$ and $\xi_n$,
\begin{equation}
    C_{SP} = (\lambda_p \xi_p + \lambda_n \xi_n)\times10^{13}{\rm cm}^{-1}\,,
    \label{CspXi}
\end{equation}
where the dimensionless coefficients $\lambda_{p,n}$ are
\begin{equation}
\begin{aligned}
    \lambda_{p,n} &= - \frac{e\mu_0}{a_B} \frac{20\sqrt2 \pi}{G_F}
    \frac{\Gamma(2\gamma+1)^2}{1+4\gamma}\\
    &\times\frac{R_0^2}{AZ\alpha}
    \left( \frac{a_B}{2ZR_0} \right)^{2\gamma} \tilde\lambda_{p,n}\times 10^{-13}{\rm cm}\,.
    \end{aligned}
\end{equation}
The numerical values of these coefficients may be found from the corresponding values for $\tilde\lambda_{p,n}$ listed in Table~\ref{tab:lambda-pn}. We present them in Table~\ref{Results} below.

The relation (\ref{CspXi}) may be used to derive limits on the couplings $\xi_{p,n}$ which follow from the experimental constraints (\ref{Csp-constr}), yielding
\begin{equation}
        |\xi_p | < 2.2\times 10^{-23} {\rm cm}\,,\qquad
    |\xi_n | < 3.0\times 10^{-23} {\rm cm}\,.
    \label{limits}
\end{equation}
Similar limits on $\xi_{p,n}$ obtained from the $^{180}$HfF$^+$ experiment \cite{HfF2017} are about an order of magnitude weaker,
$|\xi_p|<2.6\times 10^{-22}$cm, $|\xi_n|<4.5\times 10^{-22}$cm.

To summarize, we have presented a mean to relate the experimentally measured quantity $C_{SP}$ with the phenomenological parameters of the $CP$-odd nuclear interaction. We now proceed to express this relation in terms of the coupling constants of more fundamental $CP$-odd nuclear forces.

\subsection{Relation between \texorpdfstring{$C_{SP}$}{} and \texorpdfstring{$CP$}{}-odd pion-nucleon coupling constants} 

The tree-level pion exchange in known to give dominant contribution to the $CP$-odd internucleon interaction (\ref{H'}). The authors of Refs.~\cite{Schiffmoment1,FDK} established the leading-order dependence of the constants $\xi_p$ and $\xi_n$ on the $CP$-odd pion-nucleon couplings $\bar g^{(0)}_{\pi NN}$, $\bar g^{(1)}_{\pi NN}$ and $\bar g^{(2)}_{\pi NN}$ as
\begin{equation}
\label{xip-g-relation}
\begin{aligned}
    \xi_p &= -\xi_n \\
    &= 10^{-14} g (\bar g^{(1)}_{\pi NN} + 0.4 \bar g^{(2)}_{\pi NN} - 0.2 \bar g^{(0)}_{\pi NN})\,{\rm cm}\,,
\end{aligned}
\end{equation}
where $g\approx13.6$ is the strong interaction constant. Substituting this relation into Eq.~(\ref{CspXi}) allows us to find the leading-order relation between the constant $C_{SP}$ and $CP$-violating pion-nucleon couplings
\begin{equation}
    C_{SP}=\lambda_0\bar{g}^{(0)}_{\pi NN}+\lambda_1\bar{g}^{(1)}_{\pi NN}+\lambda_2\bar{g}^{(2)}_{\pi NN}\,,
    \label{2}
\end{equation}
where the numerical values of the coefficients $\lambda_0 = -0.272 (\lambda_p - \lambda_n)$, $\lambda_1 = 1.36 (\lambda_p - \lambda_n)$ and $\lambda_2 = 0.544 (\lambda_p - \lambda_n)$
are collected in Table~\ref{Results} for various paramagnetic atoms.

\subsection{Relation between \texorpdfstring{$C_{SP}$}{} and quark chromo-EDM}

In this subsection, we consider the chromo-EDM of up and down quarks denoted by $\tilde d_u$ and $\tilde d_d$, respectively. Assuming that these quantities are the only sources of $CP$-violating internucleon forces, the authors of Refs.~\cite{PospelovRitz99,FDK,PospelovRitz2001,POSPELOV2005} established the following relations:
\begin{subequations}\label{conversion}
\begin{align}
%d_p       &=1.1e(0.5\tilde{d}_d+\tilde{d}_u)\,,\\
%d_n       &=1.1e(\tilde{d}_d+0.5\tilde{d}_u)\,,\\
g\bar{g}^{(0)}_{\pi NN}&=0.8\times 10^{15}(\tilde{d}_u+\tilde{d}_d){\rm cm}^{-1}\,,\\
g\bar{g}^{(1)}_{\pi NN}&=4.0\times 10^{15}(\tilde{d}_u-\tilde{d}_d){\rm cm}^{-1}\,.
\end{align}
\end{subequations}
We substitute these relations into Eqs.~(\ref{xip-g-relation}) and (\ref{2}) and ignore the last term $\propto \bar{g}^{(2)}_{\pi NN}$ because its relation to the quark chromo-EDMs is not known. As a result, we find the leading-order dependence of $C_{SP}$ on $\tilde d_{u}$ and $\tilde d_p$,
\begin{equation}
    C_{SP}=(\lambda_u\tilde{d}_u+\lambda_d\tilde{d}_d)\times 10^{14}{\rm cm}^{-1}\,.\label{3}
\end{equation}
Numerical values of the coefficients $\lambda_u=3.84(\lambda_p-\lambda_n)$ and $\lambda_d=-4.16(\lambda_p-\lambda_n)$ are given in Table~\ref{Results} below.

\subsection{Relation between \texorpdfstring{$C_{SP}$}{} and QCD vacuum angle}

The pion-nucleon coupling constants $\bar g^{(0)}_{\pi NN}$ and $\bar g^{(1)}_{\pi NN}$ may be expressed via the QCD vacuum angle $\bar\theta$ as (see, e.g., Refs.~\cite{POSPELOV2005,deVries2015,Bsaisou2012,yamanaka2017})
\begin{subequations}\label{gandtheta}
\begin{align}
 \bar{g}^{(0)}_{\pi NN}&=-15.5\times 10^{-3}\,\bar{\theta}\,,\\
\bar{g}^{(1)}_{\pi NN}&=3.4\times 10^{-3}\,\bar{\theta}\,.   
\end{align}
\end{subequations}

Substituting these relations into Eq.~(\ref{2}), we may represent $C_{SP}$ in terms of $\bar\theta$ as
\begin{equation}
    C_{SP}=\lambda_{\theta}\times10^{-2}\bar{\theta}\label{4}\,,
\end{equation}
where the value of the constant $\lambda_{\theta}=0.88(\lambda_p-\lambda_n)$ is given in Table~\ref{Results} for different atoms. In particular, with the use of the corresponding value for $^{232}$Th, the experimental constraint on $C_{SP}$ (\ref{Csp-constr}) implies 
\begin{equation}\label{theta-constr}
    |\bar\theta| < 9\times 10^{-8}\,.
\end{equation}

This constraint is almost three orders of magnitude weaker than the currently accepted one \cite{PDG} which is based on the neutron EDM \cite{Abel2020} and Hg EDM \cite{Graner2016} experiments. However, it is comparable to the constraint on QCD vacuum angle originating from the $^{129}$Xe EDM experiments \cite{Xe1,Xe2}
 and from constraints on the nucleon EDM  \cite{FPRS,FST2020} derived from the  experiments with paramagnetic molecules. This demonstrates the importance of contributions from the $P,T$-odd nuclear forces to the atomic EDM. For reference, we collect the limits on $\xi_{p,n}$, $\bar{g}^{(0,1,2)}_{\pi NN}$, $\tilde{d}_{u,d}$ and $\bar{\theta}$ in Table \ref{ResultsLim}.
\begin{widetext}
\begin{center}
\begin{table}[htb]
\begin{tabular}{|c|c|c|c|c|c|c|c|c|c|}
\hline
\multicolumn{2}{|c|}{} & $\lambda_p$ & $\lambda_n$ & $\lambda_0$ & $\lambda_1$ & $\lambda_2$ & $\lambda_u$ & $\lambda_d$ & $\lambda_\theta$  \\
 \hline
\multirow{3}{*}{Spherical}       & ${}_{56}^{138}{\rm Ba}$ & -4.1 & -4.3 & -0.053 & 0.26 & 0.11 & 0.74 & -0.80 & 0.17 \\\cline{2-10}
                                                             & ${}_{82}^{206}{\rm Pb}$ & -3.4 & -2.8 & 0.17 & -0.83 & -0.33 & -2.3 & 2.5 & -0.54 \\ \cline{2-10}
                                                             & ${}_{82}^{208}{\rm Pb}$ & -3.4 & -2.3 & 0.29 & -1.5 & -0.58 & -4.1 & 4.5 & -0.95\\ \hline
\multirow{7}{*}[-2.5ex]{Deformed} & ${}_{70}^{172}{\rm Yb}$ & -5.6 & -3.7 & 0.50 & -2.5 & -1.0 & -7.1 & 7.7 & -1.6 \\[0.25ex] \cline{2-10}
                                                             & ${}_{70}^{174}{\rm Yb}$ & -5.6 & -4.5 & 0.31 & -1.6 & -0.62 & -4.4 & 4.8 & -1.0 \\ \cline{2-10}
                                                             & ${}_{70}^{176}{\rm Yb}$ & -5.5 & -3.5 & 0.56 & -2.8 & -1.1 & -7.9 & 8.6 & -1.8 \\ \cline{2-10}
                                                             & ${}_{72}^{178}{\rm Hf}$ & -3.9 & -2.2 & 0.47 & -2.4 & -0.94 & -6.7 & 7.2 & -1.5\\ \cline{2-10}
                                                             & ${}_{72}^{180}{\rm Hf}$ & -6.9 & -4.0 & 0.77 & -3.8 & -1.5 & -11 & -12 & -2.5\\ \cline{2-10}
                                                             & ${}_{88}^{226}{\rm Ra}$ & -2.9 & -1.8 & 0.29 & -1.4 & 0.58 & -4.1 & 4.4 & -0.94\\ \cline{2-10}
                                                             & ${}_{90}^{232}{\rm Th}$ & -3.3 & -2.4 & 0.25 & -1.2 & -0.50 & -3.5 & 3.8 & -0.81 \\ \hline
\end{tabular}
\caption{\label{Results}
The results of numerical computations of $\lambda$-coefficients in Eqs.~\eqref{CspXi}, (\ref2), (\ref3) and (\ref4).}
\end{table}
\end{center}
\end{widetext}

\begin{table}[htb]
    \centering
    \begin{tabular}{|c|c|c|c|c|c|c|c|}
    \hline
       $\frac{|\xi_{p}|}{10^{-23}{\rm cm}}$ & $\frac{|\xi_{n}|}{10^{-23}{\rm cm}}$ & $\frac{\bar{g}^{(0)}_{\pi NN}}{10^{-9}}$ & $\frac{\bar{g}^{(1)}_{\pi NN}}{10^{-9}}$ & $\frac{\bar{g}^{(2)}_{\pi NN}}{10^{-9}} $ & $\frac{\tilde{d}_u}{10^{-24}{\rm cm}}$ & $\frac{\tilde{d}_d}{10^{-24}{\rm cm}}$ & $\frac{\bar\theta}{10^{-8}}$\\ \hline
       $2.2 $ & $3.0 $ & $2.9$ & $0.6$ & $1.5$ &  $2.1$ &  $1.9$ & $9$ \\ \hline
    \end{tabular}
    \caption{Limits on $\xi_{p,n}$, $\bar{g}^{(0,1,2)}_{\pi NN}$, $\tilde{d}_{u,d}$ and $\bar\theta$ obtained from the ThO limit on $|C_{SP}|<7.3\times 10^{-10}$.}
    \label{ResultsLim}
\end{table}

\section{Conclusions}\label{Conclusion}

In this paper, we demonstrated that the eEDM experiments with paramagnetic molecules are also sensitive to the $P,T$-violating nuclear forces, as well as to other sources of $CP$ violation in the hadronic sector. We considered $P,T$-violating internucleon interaction described by the Hamiltonian (\ref{Hodd}) with coupling constants $\xi_{p,n}$. We established the leading-order relation (\ref{CspXi}) between these couplings and the constant $C_{SP}$ of the contact semileptonic interaction (\ref{contact-interaction}). This relation contains atom-specific coefficients $\lambda_{p,n}$ calculated numerically and presented in Table~\ref{Results}. We used this relation to place independent limits (\ref{limits}) on the coupling constants $\xi_{p,n}$ originating from the experimental constraint (\ref{Csp-constr}). 

The $P,T$-odd nuclear interaction is described by the phenomenological Hamiltonian which may originate from different fundamental interactions. In particular, it is known \cite{Schiffmoment1} that this interaction may appear due to the $\pi$ meson exchange between nucleons. In this case, the parameters $\xi_{p,n}$ may be expressed via the $CP$-odd pion-nucleon couplings as in Eq.~(\ref{xip-g-relation}). Then, using the known relations between the pion couplings and the quark-chromo EDMs (\ref{conversion}) as well as QCD $\bar\theta$ angle (\ref{gandtheta}) we relate the parameters $\xi_{p,n}$ to $\tilde d_{u,d}$ and to $\bar\theta$ as in Eqs.~(\ref{3}) and (\ref{4}), respectively. This allows us to place limits on $\tilde d_{u,d}$ and $\bar\theta$ originating from the experimental constraint (\ref{Csp-constr}).

We stress that the limits on $\bar g_{\pi NN}^{(0,1,2)}$, $\tilde d_{u,d}$ and $\bar\theta$ obtained in this paper are independent from similar constraints established in our recent paper \cite{FST2020} because the latter were found by taking into account nucleon EDM while in this paper we consider the $P,T$-odd nuclear force as the origin for these relations. It is useful to compare these two results. For example, the limit on the $CP$-odd pion-nucleon coupling constant $\bar g_{\pi NN}^{(2)}$ is obtained in the present paper only, the limits on $u$-quark chromo-EDM $\tilde{d}_{u}$ are approximately the same in both papers while the constraint on $\bar{\theta}$ in Eq.~(\ref{theta-constr}) is six times weaker than the analogous one found in Ref.~\cite{FST2020}, but it is comparable to the constraint on QCD vacuum angle originating from $^{129}$Xe EDM experiment \cite{Xe1,Xe2}. Of course, all these constraints on $\bar{\theta}$ are not competitive with currently accepted strongest limit $|\bar\theta|<10^{-10}$ \cite{PDG}, obtained on the neutron EDM \cite{Abel2020} and Hg EDM \cite{Graner2016}  experiments (see also discussion in Ref.~\cite{FD2020}).

It is pertinent to make some comments about the accuracy of our results. All our results are based on numerical calculations of the coefficients (\ref{lambda-coefs}) listed in Table~\ref{tab:lambda-pn}. These coefficients involve both nuclear and electronic matrix elements as well as nuclear excitation energies. The electronic matrix elements in Eq.~(\ref{M(E)}) contain radial integrals which are calculated numerically with some details given in Appendix~\ref{AppB3}. To estimate the accuracy of our numerical integration methods, we calculated similar radial integrals which are responsible for atomic energy level shifts (contributions  to  the  Lamb  shifts) due to nuclear polarizability and compared these shifts with calculated earlier values in Refs.~\cite{Plunien91,Plunien95}. This allows us to conclude that the error in numerical calculation of electronic matrix elements does not exceed 5\%. However, the main source of uncertainty is represented by nuclear matrix elements and nuclear excitation energies. In this paper, we use the single-particle nuclear shell model to estimate M1 spin-flip matrix elements and corresponding energies in heavy nuclei listed in Appendix~\ref{NuclearSection}. Using these quantities we calculated reduced transition probabilities for M1 spin-flip transitions for some nuclei (ytterbium and thorium). The results have been compared  with analogous quantities calculated using sophisticated many-body methods in Ref.~\cite{nuc-res}. On the basis of this comparison we estimate the error in nuclear transition energies and matrix elements to be under 45-50\%. %As a result, we conclude that the computational errors of coefficients in Table~\ref{Results} do not exceed 50\%. We stress that this level of accuracy is typical for most of calculations of hadron $CP$-violating effects in heavy elements and is acceptable withing the goals of this paper.

To conclude, we stress that the EDM-like experiments with molecules in paramagnetic state are sensitive to hadronic $CP$-violating parameters such as $CP$-odd pion-nucleon couplings, quark-chromo EDMs and QCD vacuum angle. We expect that results from the next generation of these experiments will significantly improve the limits on these parameters.

\section*{Acknowledgements}
This work was supported by the Australian Research Council Grants No. DP190100974 and DP200100150 and the Gutenberg Fellowship. 

\appendix
\section{Nuclear energies and matrix elements}\label{NuclearSection}

These appendices follow our calculation in Ref. \cite{FST2020} and are presented here to provide the reader with the details facilitating the understanding of the current paper. In the first appendix we  estimate the matrix elements and corresponding energies of the nuclear M1 spin-flip single-particle transitions. The details of these computations slightly differ for (nearly) spherical and deformed nuclei. Therefore, we consider these two cases separately.

\subsection{Spherical nuclei}

In this section, we focus on the $^{208}$Pb, $^{206}$Pb and $^{138}$Ba nuclei, which are nearly spherical, i.e., they have deformation $\delta<0.1$. For these nuclei, proton and neutron single-particle states may be labeled as $|n,l,j,m\rangle$, where $n$ is the oscillator quantum number, $l$ and $j$ are the orbital and total momentum numbers, $m$ is magnetic quantum number. In this basis, the nuclear spin operator $\bf s$ provides transitions between fine structure doublets. 

In the $^{208}$Pb nucleus, the non-vanishing matrix elements of the spin operator are
$\langle 5h\frac92 |{\bf s} |5h\frac{11}{2}\rangle$ for protons and $\langle 6i\frac{11}{2} |{\bf s} |6i\frac{13}{2}\rangle$ for neutrons. The isotope $^{206}$Pb receives additional contributions from the $\langle 5p\frac12 |{\bf s}|5p\frac32\rangle$ neutron matrix elements. For $^{138}$Ba, non-vanishing proton contributions arise from the matrix elements $\langle 4d\frac32|{\bf s}|4d\frac52 \rangle$ and $\langle 4g\frac92|{\bf s}|4g\frac72 \rangle$ whereas neutron contributions come from $\langle 5h\frac92 | {\bf s}| 5h\frac{11}2\rangle$. All these matrix elements may be calculated using the properties of spherical spinors (see, e.g., Ref.\ \cite{Recurrence}). The energies of all these transitions may be estimated with the use of Fig.\ 5 in Ref.\ \cite{BM}. When the energies are (nearly) degenerate, we give the sum of matrix elements corresponding to the same energy.
In the Table \ref{sphericalNuc} below, we collect the values of such matrix elements with the corresponding energies for $^{208}$Pb, $^{206}$Pb and $^{138}$Ba. The values for the nuclear radii $R_0$ and the energy of giant dipole resonance are calculated according to the empirical formulas:
\begin{eqnarray}
R_0 &=& 1.2 A^{1/3}\, {\rm fm}\,,\\
\Delta \bar E &=& 95 A^{-1/3}(1 - A^{-1/3})\, {\rm MeV}\,.
\label{DeltaBarE}
\end{eqnarray}
For reference, the values of the deformation parameter $\delta$ are also presented.

\subsection{Deformed nuclei}

For deformed heavy nuclei with $\delta>0.1$, it is convenient to use the Nilsson basis \cite{Nilsson,BM}, wherein proton and neutron single-particle states are labeled with $|n_3,n_\perp,\Lambda,\Omega\rangle$, where $n_3$ and $n_\perp$ are the oscillator quantum numbers, $\Lambda$ and $\Omega$ are the projections of angular and total momenta on the deformation axis. Note that $\Omega = \Lambda + \Sigma$ where $\Sigma$ is the projection of the nucleon's spin on the deformation axis. The dependence of the energy levels on the deformation parameter $\delta$ in this model may be inferred from Fig.\ 5 in Ref.\ \cite{BM}. From such dependence, one may estimate the energies of the spin-flip transitions. Note that in the basis $|n_3,n_\perp,\Lambda,\Omega\rangle$, each M1 spin-flip matrix element is $\bra{m'} s_+ \ket{0'} =1$, and the corresponding energy level is doubly degenerate since each quantum number $\Sigma$ corresponds to $\pm \Lambda$. 

The single-nucleon spin-flip transition energies $\Delta E_{n'}$, the energies the of giant dipole resonance $\Delta \bar E$, the deformation parameters $\delta$ and the nuclear radii $R_0$ for several nuclei of interest are presented in Table \ref{deformNuc}.

%%%%%%%%%%%%%%%%%%%%%%%%%%%%%%%%%%%%%%%%%%%%%%%%%%%%%%%%%%%%%%%%%%%%%%%%%%%%%%%%%%%%%%%%%%%%

\section{Evaluation of electronic matrix elements}
\label{AppB}

In this appendix, we provide the details for the numerical calculation of the electronic matrix element \eqref{M(E)}. For convenience, we use the spherical basis $({\bf e}_+, {\bf e}_-, {\bf e}_0)$. The components of the vectors in this basis will be labeled by the $(+,-,0)$ subscripts. Due to spherical symmetry, Eq.~\eqref{M(E)} may be rewritten in terms of the `0'-component of the operators $\hat{\bf R}$ and $\hat{\bf R}\times \boldsymbol{\alpha}$
\begin{widetext}\begin{equation}
%\begin{aligned}
M(\Delta E_{n'})
=\frac{\alpha}{c_{s_{1/2}}c_{p_{1/2}}}
\sum_{n}\frac{\bra{p_{1/2}}f(R)\hat{\bf R}_0\ket{n}\bra{n}f(R)(\hat{\bf R}\times\boldsymbol{\alpha})_0\ket{s_{1/2}}}{\Delta E_{n}+{\rm sgn}(E_n)\Delta E_{n'}} + (s_{1/2}\leftrightarrow p_{1/2})\,.
%\\
%&+\frac{\alpha}{c_{s_{1/2}}c_{p_{1/2}}}
%\sum_{n}\frac{\bra{s_{1/2}}M_0\ket{n}\bra{n}D_0\ket{p_{1/2}}}{E_{s_{1/2}}-E_{n}-{\rm sgn}(E_n)\Delta E_{n'}}\,.
\label{0}
%\end{aligned}
\end{equation}
\end{widetext}

For further computation of the matrix elements in Eq.~\eqref{0} the electron wave functions need to be specified.

\subsection{The \texorpdfstring{$s_{1/2}$}{} and \texorpdfstring{$p_{1/2}$}{} wave functions}
\label{DiracCoilombF}
The valence electron $s_{1/2}$ and $p_{1/2}$ wave functions may be expressed in terms of the spherical spinors $\Omega^\kappa_\mu(\hat{\bf R})$ where $\mu$ is the magnetic quantum number and $\kappa=(l-j)(2j+1)$ as
\begin{subequations}
\label{B2}
\begin{align}
\ket{s_{1/2}}&=c_{s_{1/2}}\left(
\begin{array}{c}
f_{s_{1/2}}(R)\Omega^{-1}_{\mu }(\hat{\bf R}) \\
ig_{s_{1/2}}(R)\Omega^{1}_{\mu}(\hat{\bf R})
\end{array}\right)\,,\\
\ket{p_{1/2}}&=c_{p_{1/2}}\left(
\begin{array}{c}
f_{p_{1/2}}(R)\Omega^{1}_{\mu }(\hat{\bf R}) \\
ig_{p_{1/2}}(R)\Omega^{-1}_{\mu}(\hat{\bf R})
\end{array}
\right)\,,
\end{align} 
\end{subequations}
where the radial wave functions $f_{s,p_{1/2}}$ and $g_{s,p_{1/2}}$ are well approximated in the region $R_0<R\ll a_B/Z^{1/3}$ by the Bessel functions of the first kind $J_\nu(x)$ (see, e.g., \cite{khriplovich1991parity}),
\begin{subequations}\label{Bessel}
\begin{align}
f_{s_{1/2}}(R)&=\frac{(-1+\gamma)J_{2\gamma}(x)- \frac{x}{2}J_{2\gamma-1}(x)}{R}\,,\\
f_{p_{1/2}}(R)&=\frac{( 1+\gamma)J_{2\gamma}(x)- \frac{x}{2}J_{2\gamma-1}(x)}{R}\,,\\
g_{s_{1/2}}(R)&=g_{p_{1/2}}(R)=\frac{Z\alpha J_{2\gamma}(x)}{R} \,,
\end{align}
\end{subequations}
where $x\equiv\sqrt{8ZR/a_B}$.

Note that the wave functions (\ref{Bessel}) are the zero-energy solutions of the Dirac-Coulomb equations for a point-like nucleus. For an extended nucleus, the corresponding solution is complicated. At the current level of accuracy, it suffices to use Eqs.\ \eqref{Bessel} as an approximation to the wave functions. For the region inside the nucleus, $0\leq R\leq R_0$, the radial wave functions $f_{s,p_{1/2}}$ and $g_{s,p_{1/2}}$ may be continued as follows
\begin{subequations}\label{Bessel-inside}
\begin{align}
f_{s_{1/2}}(R)&=\frac{(-1+\gamma)J_{2\gamma}(x_0)- \frac{x_0}{2}J_{2\gamma-1}(x_0)}{R_0}\,,\\
f_{p_{1/2}}(R)&=\frac{R\left[( 1+\gamma)J_{2\gamma}(x_0)- \frac{x_0}{2}J_{2\gamma-1}(x_0)\right]}{R_0^2}\,,\\
g_{s_{1/2}}(R)&=\frac{R}{R_0}Z\alpha J_{2\gamma}(x_0)\,,\\
g_{p_{1/2}}(R)&=\frac{1}{R_0}Z\alpha J_{2\gamma}(x_0)\,,
\end{align}
\end{subequations}
where $x_0\equiv\sqrt{8ZR_0/a_B}$.

Note that these functions are the approximate solutions (containing only leading terms at small distance) of the Dirac equation inside the nucleus with constant density.

\subsection{Excited electronic states of the continuous spectrum}
\label{AppB2}
The excited electronic states $|n\rangle$ in the continuous spectrum may be labeled by the quantum number $\kappa=(l-j)(2j+1)$ and the energy $E$, $|n\rangle \equiv|E\kappa\rangle$. In spherical coordinates, these functions read (see, e.g., Refs.\ \cite{Landau4,greiner2000relativistic}): 
\begin{equation}
\label{B6}
\ket{n}\equiv\ket{E\kappa}=\left(\begin{matrix}
f_{\kappa}^E(R)\Omega^{\kappa}_{\mu}(\hat{\bf R})\\
ig_{\kappa}^E(R)\Omega^{-\kappa}_{\mu}(\hat{\bf R})
\end{matrix}\right)\,,
\end{equation}
with
\begin{subequations}\label{HyperGeo}
\begin{align}
&f_{\kappa}^E(R) =\frac{(2pR)^\gamma e^{\pi y/2}\left|\Gamma(\gamma+iy)\right|\sqrt{\left|E+ m_e\right|}}{R\sqrt{\pi p}\Gamma(2\gamma+1)}\nonumber\\
&\times{\rm Re}[e^{-ipR+i\eta}{}_1F_1(\gamma+1+iy,2\gamma+1,2ipR)]\,,\label{fCont}\\
&g_{\kappa}^E(R) =-{\rm sgn}(E)\frac{(2pR)^\gamma e^{\pi y/2}\left|\Gamma(\gamma+iy)\right|\sqrt{\left|E- m_e\right|}}{r\sqrt{\pi p}\Gamma(2\gamma+1)}\nonumber\\
&\times{\rm Im}[e^{-ipR+i\eta}{}_1F_1(\gamma+1+iy,2\gamma+1,2ipR)]\,.\label{gCont}
\end{align}
\end{subequations}
Here $p=\sqrt{E^2-m_e^2}$ is the electron's momentum, $y=Z\alpha E/p$, $e^{i\eta}=\sqrt{-\frac{\kappa-iym_e/E}{\gamma+iy}}$ and $_1F_1(a,b,z)$ is the confluent hypergeometric function of the first kind. Note that the wave functions (\ref{B6}) are normalized as $\braket{E'\kappa|E\kappa}=\delta(E'-E)$.

The functions (\ref{HyperGeo}) solve for the Dirac equation with a point-like nucleus. Therefore, we will only use them for outside of the nucleus, $R>R_0$. For the inside of the nucleus, $0\leq R\leq R_0$, we will consider the following continuation of these functions
\begin{equation}\label{Inside}
f^E_\kappa(R)=b_1 R^l\,,\qquad g^E_\kappa(R)=b_2 R^{\tilde{l}}\,,
\end{equation}
where $l=|\kappa+1/2|-1/2$ is the orbital angular momentum corresponding to $\kappa$, $\tilde{l}=|-\kappa+1/2|-1/2$ is the orbital angular momentum corresponding to $-\kappa$. The values of the coefficients $b_{1}$ and $b_2$ are determined by matching Eqs.~(\ref{HyperGeo}) and (\ref{Inside}) on the boundary of the nucleus. The wave functions \eqref{Inside} are, to the leading order, solutions to the Dirac equation inside a nucleus of a constant density.

We stress that the extension of the electronic wave functions to the inside region of the nucleus (\ref{Inside}) is an approximation which is acceptable at our level of accuracy. We checked the validity of this approximation by computing the Lamb shift in heavy atoms due to nuclear polarizability. Within this approximation, we have 95\% agreement with the exact results presented in Refs.~\cite{Plunien91,Pachucki93,Plunien95}.

\subsection{Results of calculation of electronic matrix element}
\label{AppB3}
Substituting the wave functions (\ref{B2}) and (\ref{B6}) into Eq.~(\ref0) and performing the integration over angular variables, we obtain
\begin{equation}
\begin{aligned}
M(\Delta E_{n'})&=-\frac{2\alpha }{9}\int\limits_{m_e}^{\infty}\frac{T(E)dE}{E-E_{s_{1/2}}+\Delta E_{n'}}\\
&-\frac{2\alpha }{9}\int\limits_{-\infty }^{-m_e}\frac{T(E)dE}{E-E_{s_{1/2}}-\Delta E_{n'}}\,,
\label{EnergyInt}    
\end{aligned}
\end{equation}
where
\begin{equation}
\begin{aligned}
   T(E&)=R_s^1(E)R_p^1(E)-R_s^{-2}(E)R_p^{-2}(E)\\
 &-S_s^{-1}(E)S_p^{-1}(E)+S_s^2(E)S_p^2(E)\,, \label{T(E)}
\end{aligned}
\end{equation}
and the radial integrals $R_{s,p}^\kappa(E)$ and $S_{s,p}^\kappa(E)$ are defined by
\begin{subequations}
\label{RadialIntegrals}
\begin{align}
 R_s^\kappa(E)&\equiv \int_0^\infty{\left(f_{s_{1/2}}f_{\kappa}^E+ g_{s_{1/2}}g_{\kappa}^E \right)f( R){{R}^{2}}dR} \,,\\
%\end{align}
%\begin{align}
 R_p^\kappa(E)&\equiv \int_0^\infty{\left(f_{p_{1/2}}g_{\kappa}^E+ g_{p_{1/2}}f_{\kappa}^E  \right)f( R ){{R}^{2}}dR} \,,\\
 S_s^\kappa(E)&\equiv \int_0^\infty{\left(f_{s_{1/2}}g_{\kappa}^E+ g_{s_{1/2}}f_{\kappa}^E \right)f( R ){{R}^{2}}dR} \,,\\ 
 S_p^\kappa(E)&\equiv \int_0^\infty{\left(f_{p_{1/2}}f_{\kappa}^E+ g_{p_{1/2}}g_{\kappa}^E \right)f( R ){{R}^{2}}dR} \,.
\end{align}
\end{subequations}
Here the radial function $f(R)$ is given by Eq.~(\ref{8}).
Note that Eq.~(\ref{T(E)}) involves only the terms with $\kappa=\pm 1,\pm 2$ which are allowed by the selection rules for transitions from $s_{1/2}$ and $p_{1/2}$ bound electron states.

With the radial wave functions (\ref{Bessel}), (\ref{Bessel-inside}), (\ref{HyperGeo}) and (\ref{Inside}), the radial integrals (\ref{RadialIntegrals}) may be computed numerically for any specific electron energy $E$ and nuclear energy $\Delta E_{n'}$ or $\Delta\bar E$. For all values of $\Delta E_{n'}$ and $\Delta\bar E$ presented in Appendix \ref{NuclearSection}, numerical analysis showed that for $|E|>500m_e$, $T(E)/(E_{s_{1/2}}-E\pm\Delta E_{n'})$ is effectively zero, so the energy integrals in Eqs.~\eqref{EnergyInt} may be cut off at $|E|\approx 500m_e$.  We also point out that the dominant contributions to the energy integrals \eqref{EnergyInt} come from the region where $E\sim 50m_e$, which is larger than the values of $\Delta E_{n'}$ or $\Delta\bar E$ considered in Appendix \ref{NuclearSection}. As a result, $M(\Delta E_{n'})$ is a slowly varying function of energy.

The energy integrals in Eqs.~\eqref{EnergyInt} are computed numerically, giving $M(\Delta E_{n'})$ for all values of $\Delta E_{n'}$ and $\Delta\bar E$ presented in Appendix A. The resulting numerical values of the electronic factors $M_p$, $M_n$ and $M(\Delta\bar E)$ are presented in Table~\ref{Electronic} below.

\bibliography{Bib}

\begin{thebibliography}{37}
\expandafter\ifx\csname natexlab\endcsname\relax\def\natexlab#1{#1}\fi
\expandafter\ifx\csname bibnamefont\endcsname\relax
  \def\bibnamefont#1{#1}\fi
\expandafter\ifx\csname bibfnamefont\endcsname\relax
  \def\bibfnamefont#1{#1}\fi
\expandafter\ifx\csname citenamefont\endcsname\relax
  \def\citenamefont#1{#1}\fi
\expandafter\ifx\csname url\endcsname\relax
  \def\url#1{\texttt{#1}}\fi
\expandafter\ifx\csname urlprefix\endcsname\relax\def\urlprefix{URL }\fi
\providecommand{\bibinfo}[2]{#2}
\providecommand{\eprint}[2][]{\url{#2}}

\bibitem[{\citenamefont{Purcell and Ramsey}(1950)}]{Purcell1950}
\bibinfo{author}{\bibfnamefont{E.~M.} \bibnamefont{Purcell}} \bibnamefont{and}
  \bibinfo{author}{\bibfnamefont{N.~F.} \bibnamefont{Ramsey}},
  \bibinfo{journal}{Phys. Rev.} \textbf{\bibinfo{volume}{78}},
  \bibinfo{pages}{807} (\bibinfo{year}{1950}).

\bibitem[{\citenamefont{Lee and Yang}(1957)}]{leeyang1957}
\bibinfo{author}{\bibfnamefont{T.-D.} \bibnamefont{Lee}} \bibnamefont{and}
  \bibinfo{author}{\bibfnamefont{C.~N.} \bibnamefont{Yang}},
  \bibinfo{type}{Tech. Rep.}, \bibinfo{institution}{Brookhaven National Lab.,
  Upton, NY} (\bibinfo{year}{1957}).

\bibitem[{\citenamefont{Landau}(1957{\natexlab{a}})}]{LANDAU57a}
\bibinfo{author}{\bibfnamefont{L.}~\bibnamefont{Landau}},
  \bibinfo{journal}{Sov. Phys. - JETP} \textbf{\bibinfo{volume}{5}},
  \bibinfo{pages}{336} (\bibinfo{year}{1957}{\natexlab{a}}).

\bibitem[{\citenamefont{Landau}(1957{\natexlab{b}})}]{LANDAU57b}
\bibinfo{author}{\bibfnamefont{L.}~\bibnamefont{Landau}},
  \bibinfo{journal}{Nuclear Physics} \textbf{\bibinfo{volume}{3}},
  \bibinfo{pages}{127 } (\bibinfo{year}{1957}{\natexlab{b}}), ISSN
  \bibinfo{issn}{0029-5582}.

\bibitem[{\citenamefont{Sakharov}(1991)}]{Sakharov91}
\bibinfo{author}{\bibfnamefont{A.~D.} \bibnamefont{Sakharov}},
  \bibinfo{journal}{Sov. Phys. Usp.} \textbf{\bibinfo{volume}{34}},
  \bibinfo{pages}{392} (\bibinfo{year}{1991}).

\bibitem[{\citenamefont{Andreev et~al.}(2018)\citenamefont{Andreev, Ang,
  DeMille, Doyle, Gabrielse, Haefner, Hutzler, Lasner, Meisenhelder, O’Leary
  et~al.}}]{ACMEII}
\bibinfo{author}{\bibfnamefont{V.}~\bibnamefont{Andreev}},
  \bibinfo{author}{\bibfnamefont{D.~G.} \bibnamefont{Ang}},
  \bibinfo{author}{\bibfnamefont{D.}~\bibnamefont{DeMille}},
  \bibinfo{author}{\bibfnamefont{J.~M.} \bibnamefont{Doyle}},
  \bibinfo{author}{\bibfnamefont{G.}~\bibnamefont{Gabrielse}},
  \bibinfo{author}{\bibfnamefont{J.}~\bibnamefont{Haefner}},
  \bibinfo{author}{\bibfnamefont{N.~R.} \bibnamefont{Hutzler}},
  \bibinfo{author}{\bibfnamefont{Z.}~\bibnamefont{Lasner}},
  \bibinfo{author}{\bibfnamefont{C.}~\bibnamefont{Meisenhelder}},
  \bibinfo{author}{\bibfnamefont{B.~R.} \bibnamefont{O’Leary}},
  \bibnamefont{et~al.} (\bibinfo{collaboration}{ACME}),
  \bibinfo{journal}{Nature} \textbf{\bibinfo{volume}{562}},
  \bibinfo{pages}{355} (\bibinfo{year}{2018}).

\bibitem[{\citenamefont{Loh et~al.}(2013)\citenamefont{Loh, Cossel, Grau, Ni,
  Meyer, Bohn, Ye, and Cornell}}]{Loh2013}
\bibinfo{author}{\bibfnamefont{H.}~\bibnamefont{Loh}},
  \bibinfo{author}{\bibfnamefont{K.~C.} \bibnamefont{Cossel}},
  \bibinfo{author}{\bibfnamefont{M.~C.} \bibnamefont{Grau}},
  \bibinfo{author}{\bibfnamefont{K.-K.} \bibnamefont{Ni}},
  \bibinfo{author}{\bibfnamefont{E.~R.} \bibnamefont{Meyer}},
  \bibinfo{author}{\bibfnamefont{J.~L.} \bibnamefont{Bohn}},
  \bibinfo{author}{\bibfnamefont{J.}~\bibnamefont{Ye}}, \bibnamefont{and}
  \bibinfo{author}{\bibfnamefont{E.~A.} \bibnamefont{Cornell}},
  \bibinfo{journal}{Science} \textbf{\bibinfo{volume}{342}},
  \bibinfo{pages}{1220} (\bibinfo{year}{2013}).

\bibitem[{\citenamefont{Yamanaka et~al.}(2017)\citenamefont{Yamanaka, Sahoo,
  Yoshinaga, Sato, Asahi, and Das}}]{yamanaka2017}
\bibinfo{author}{\bibfnamefont{N.}~\bibnamefont{Yamanaka}},
  \bibinfo{author}{\bibfnamefont{B.}~\bibnamefont{Sahoo}},
  \bibinfo{author}{\bibfnamefont{N.}~\bibnamefont{Yoshinaga}},
  \bibinfo{author}{\bibfnamefont{T.}~\bibnamefont{Sato}},
  \bibinfo{author}{\bibfnamefont{K.}~\bibnamefont{Asahi}}, \bibnamefont{and}
  \bibinfo{author}{\bibfnamefont{B.}~\bibnamefont{Das}}, \bibinfo{journal}{Eur.
  Phys. J. A} \textbf{\bibinfo{volume}{53}}, \bibinfo{pages}{54}
  (\bibinfo{year}{2017}).

\bibitem[{\citenamefont{Chupp et~al.}(2019)\citenamefont{Chupp, Fierlinger,
  Ramsey-Musolf, and Singh}}]{chupp2019}
\bibinfo{author}{\bibfnamefont{T.~E.} \bibnamefont{Chupp}},
  \bibinfo{author}{\bibfnamefont{P.}~\bibnamefont{Fierlinger}},
  \bibinfo{author}{\bibfnamefont{M.~J.} \bibnamefont{Ramsey-Musolf}},
  \bibnamefont{and} \bibinfo{author}{\bibfnamefont{J.~T.} \bibnamefont{Singh}},
  \bibinfo{journal}{Rev. Mod. Phys.} \textbf{\bibinfo{volume}{91}},
  \bibinfo{pages}{015001} (\bibinfo{year}{2019}).

\bibitem[{\citenamefont{Flambaum
  et~al.}(2020{\natexlab{a}})\citenamefont{Flambaum, Pospelov, Ritz, and
  Stadnik}}]{FPRS}
\bibinfo{author}{\bibfnamefont{V.~V.} \bibnamefont{Flambaum}},
  \bibinfo{author}{\bibfnamefont{M.}~\bibnamefont{Pospelov}},
  \bibinfo{author}{\bibfnamefont{A.}~\bibnamefont{Ritz}}, \bibnamefont{and}
  \bibinfo{author}{\bibfnamefont{Y.~V.} \bibnamefont{Stadnik}},
  \bibinfo{journal}{Phys. Rev. D} \textbf{\bibinfo{volume}{102}},
  \bibinfo{pages}{035001} (\bibinfo{year}{2020}{\natexlab{a}}).

\bibitem[{\citenamefont{Flambaum
  et~al.}(2020{\natexlab{b}})\citenamefont{Flambaum, Samsonov, and {Tran
  Tan}}}]{FST2020}
\bibinfo{author}{\bibfnamefont{V.~V.} \bibnamefont{Flambaum}},
  \bibinfo{author}{\bibfnamefont{I.~B.} \bibnamefont{Samsonov}},
  \bibnamefont{and} \bibinfo{author}{\bibfnamefont{H.~B.} \bibnamefont{{Tran
  Tan}}} (\bibinfo{year}{2020}{\natexlab{b}}), \eprint{arXiv:2004.10359}.

\bibitem[{\citenamefont{Khriplovich}(1991)}]{khriplovich1991parity}
\bibinfo{author}{\bibfnamefont{I.~B.} \bibnamefont{Khriplovich}},
  \emph{\bibinfo{title}{Parity nonconservation in atomic phenomena}}
  (\bibinfo{publisher}{Gordon and Breach Science Publishers},
  \bibinfo{year}{1991}).

\bibitem[{\citenamefont{Sushkov et~al.}(1984)\citenamefont{Sushkov, Flambaum,
  and Khriplovich}}]{Schiffmoment1}
\bibinfo{author}{\bibfnamefont{O.~P.} \bibnamefont{Sushkov}},
  \bibinfo{author}{\bibfnamefont{V.~V.} \bibnamefont{Flambaum}},
  \bibnamefont{and} \bibinfo{author}{\bibfnamefont{I.~B.}
  \bibnamefont{Khriplovich}}, \bibinfo{journal}{Zh. Eksp. Teor. Fiz}
  \textbf{\bibinfo{volume}{87}}, \bibinfo{pages}{1521} (\bibinfo{year}{1984}).

\bibitem[{\citenamefont{Bsaisou et~al.}(2015)\citenamefont{Bsaisou, Meißner,
  Nogga, and Wirzba}}]{Bsaisou}
\bibinfo{author}{\bibfnamefont{J.}~\bibnamefont{Bsaisou}},
  \bibinfo{author}{\bibfnamefont{U.-G.} \bibnamefont{Meißner}},
  \bibinfo{author}{\bibfnamefont{A.}~\bibnamefont{Nogga}}, \bibnamefont{and}
  \bibinfo{author}{\bibfnamefont{A.}~\bibnamefont{Wirzba}},
  \bibinfo{journal}{Annals Phys.} \textbf{\bibinfo{volume}{359}},
  \bibinfo{pages}{317} (\bibinfo{year}{2015}).

\bibitem[{\citenamefont{Flambaum and Dzuba}(2020)}]{FD2020}
\bibinfo{author}{\bibfnamefont{V.~V.} \bibnamefont{Flambaum}} \bibnamefont{and}
  \bibinfo{author}{\bibfnamefont{V.~A.} \bibnamefont{Dzuba}},
  \bibinfo{journal}{Phys. Rev. A} \textbf{\bibinfo{volume}{101}},
  \bibinfo{pages}{042504} (\bibinfo{year}{2020}).

\bibitem[{\citenamefont{Bohr and Mottelson}(1998)}]{BM}
\bibinfo{author}{\bibfnamefont{A.}~\bibnamefont{Bohr}} \bibnamefont{and}
  \bibinfo{author}{\bibfnamefont{B.~R.} \bibnamefont{Mottelson}},
  \emph{\bibinfo{title}{Nuclear Structure}}, vol.~\bibinfo{volume}{2}
  (\bibinfo{publisher}{World Scientific}, \bibinfo{address}{Singapore},
  \bibinfo{year}{1998}).

\bibitem[{\citenamefont{Plunien et~al.}(1991)\citenamefont{Plunien, M\"uller,
  Greiner, and Soff}}]{Plunien91}
\bibinfo{author}{\bibfnamefont{G.}~\bibnamefont{Plunien}},
  \bibinfo{author}{\bibfnamefont{B.}~\bibnamefont{M\"uller}},
  \bibinfo{author}{\bibfnamefont{W.}~\bibnamefont{Greiner}}, \bibnamefont{and}
  \bibinfo{author}{\bibfnamefont{G.}~\bibnamefont{Soff}},
  \bibinfo{journal}{Phys. Rev. A} \textbf{\bibinfo{volume}{43}},
  \bibinfo{pages}{5853} (\bibinfo{year}{1991}).

\bibitem[{\citenamefont{Plunien and Soff}(1995)}]{Plunien95}
\bibinfo{author}{\bibfnamefont{G.}~\bibnamefont{Plunien}} \bibnamefont{and}
  \bibinfo{author}{\bibfnamefont{G.}~\bibnamefont{Soff}},
  \bibinfo{journal}{Phys. Rev. A} \textbf{\bibinfo{volume}{51}},
  \bibinfo{pages}{1119} (\bibinfo{year}{1995}).

\bibitem[{\citenamefont{Flambaum and Khriplovich}(1985)}]{NewBounds1985}
\bibinfo{author}{\bibfnamefont{V.~V.} \bibnamefont{Flambaum}} \bibnamefont{and}
  \bibinfo{author}{\bibfnamefont{I.~B.} \bibnamefont{Khriplovich}},
  \bibinfo{journal}{Zh. Eksp. Theor. Fiz} \textbf{\bibinfo{volume}{89}},
  \bibinfo{pages}{1505} (\bibinfo{year}{1985}).

\bibitem[{\citenamefont{Cairncross et~al.}(2017)\citenamefont{Cairncross,
  Gresh, Grau, Cossel, Roussy, Ni, Zhou, Ye, and Cornell}}]{HfF2017}
\bibinfo{author}{\bibfnamefont{W.~B.} \bibnamefont{Cairncross}},
  \bibinfo{author}{\bibfnamefont{D.~N.} \bibnamefont{Gresh}},
  \bibinfo{author}{\bibfnamefont{M.}~\bibnamefont{Grau}},
  \bibinfo{author}{\bibfnamefont{K.~C.} \bibnamefont{Cossel}},
  \bibinfo{author}{\bibfnamefont{T.~S.} \bibnamefont{Roussy}},
  \bibinfo{author}{\bibfnamefont{Y.}~\bibnamefont{Ni}},
  \bibinfo{author}{\bibfnamefont{Y.}~\bibnamefont{Zhou}},
  \bibinfo{author}{\bibfnamefont{J.}~\bibnamefont{Ye}}, \bibnamefont{and}
  \bibinfo{author}{\bibfnamefont{E.~A.} \bibnamefont{Cornell}},
  \bibinfo{journal}{Phys. Rev. Lett.} \textbf{\bibinfo{volume}{119}},
  \bibinfo{pages}{153001} (\bibinfo{year}{2017}).

\bibitem[{\citenamefont{Flambaum et~al.}(2014)\citenamefont{Flambaum, DeMille,
  and Kozlov}}]{FDK}
\bibinfo{author}{\bibfnamefont{V.~V.} \bibnamefont{Flambaum}},
  \bibinfo{author}{\bibfnamefont{D.}~\bibnamefont{DeMille}}, \bibnamefont{and}
  \bibinfo{author}{\bibfnamefont{M.~G.} \bibnamefont{Kozlov}},
  \bibinfo{journal}{Phys. Rev. Lett.} \textbf{\bibinfo{volume}{113}},
  \bibinfo{pages}{103003} (\bibinfo{year}{2014}).

\bibitem[{\citenamefont{Pospelov and Ritz}(1999)}]{PospelovRitz99}
\bibinfo{author}{\bibfnamefont{M.}~\bibnamefont{Pospelov}} \bibnamefont{and}
  \bibinfo{author}{\bibfnamefont{A.}~\bibnamefont{Ritz}},
  \bibinfo{journal}{Phys. Rev. Lett.} \textbf{\bibinfo{volume}{83}},
  \bibinfo{pages}{2526} (\bibinfo{year}{1999}).

\bibitem[{\citenamefont{Pospelov and Ritz}(2001)}]{PospelovRitz2001}
\bibinfo{author}{\bibfnamefont{M.}~\bibnamefont{Pospelov}} \bibnamefont{and}
  \bibinfo{author}{\bibfnamefont{A.}~\bibnamefont{Ritz}},
  \bibinfo{journal}{Phys. Rev. D} \textbf{\bibinfo{volume}{63}},
  \bibinfo{pages}{073015} (\bibinfo{year}{2001}).

\bibitem[{\citenamefont{Pospelov and Ritz}(2005)}]{POSPELOV2005}
\bibinfo{author}{\bibfnamefont{M.}~\bibnamefont{Pospelov}} \bibnamefont{and}
  \bibinfo{author}{\bibfnamefont{A.}~\bibnamefont{Ritz}},
  \bibinfo{journal}{Ann. Phys.} \textbf{\bibinfo{volume}{318}},
  \bibinfo{pages}{119 } (\bibinfo{year}{2005}), \bibinfo{note}{special Issue}.

\bibitem[{\citenamefont{de~Vries et~al.}(2015)\citenamefont{de~Vries,
  Mereghetti, and Walker-Loud}}]{deVries2015}
\bibinfo{author}{\bibfnamefont{J.}~\bibnamefont{de~Vries}},
  \bibinfo{author}{\bibfnamefont{E.}~\bibnamefont{Mereghetti}},
  \bibnamefont{and}
  \bibinfo{author}{\bibfnamefont{A.}~\bibnamefont{Walker-Loud}},
  \bibinfo{journal}{Phys. Rev. C} \textbf{\bibinfo{volume}{92}},
  \bibinfo{pages}{045201} (\bibinfo{year}{2015}).

\bibitem[{\citenamefont{Bsaisou et~al.}(2013)\citenamefont{Bsaisou, Hanhart,
  Liebig, Meissner, Nogga, and Wirzba}}]{Bsaisou2012}
\bibinfo{author}{\bibfnamefont{J.}~\bibnamefont{Bsaisou}},
  \bibinfo{author}{\bibfnamefont{C.}~\bibnamefont{Hanhart}},
  \bibinfo{author}{\bibfnamefont{S.}~\bibnamefont{Liebig}},
  \bibinfo{author}{\bibfnamefont{U.-G.} \bibnamefont{Meissner}},
  \bibinfo{author}{\bibfnamefont{A.}~\bibnamefont{Nogga}}, \bibnamefont{and}
  \bibinfo{author}{\bibfnamefont{A.}~\bibnamefont{Wirzba}},
  \bibinfo{journal}{Eur. Phys. J. A} \textbf{\bibinfo{volume}{49}},
  \bibinfo{pages}{31} (\bibinfo{year}{2013}).

\bibitem[{\citenamefont{Tanabashi et~al.}(2018)}]{PDG}
\bibinfo{author}{\bibfnamefont{M.}~\bibnamefont{Tanabashi}}
  \bibnamefont{et~al.} (\bibinfo{collaboration}{Particle Data Group}),
  \bibinfo{journal}{Phys. Rev. D} \textbf{\bibinfo{volume}{98}},
  \bibinfo{pages}{030001} (\bibinfo{year}{2018}).

\bibitem[{\citenamefont{Abel et~al.}(2020)\citenamefont{Abel, Afach, Ayres,
  Baker, Ban, Bison, Bodek, Bondar, Burghoff, Chanel et~al.}}]{Abel2020}
\bibinfo{author}{\bibfnamefont{C.}~\bibnamefont{Abel}},
  \bibinfo{author}{\bibfnamefont{S.}~\bibnamefont{Afach}},
  \bibinfo{author}{\bibfnamefont{N.~J.} \bibnamefont{Ayres}},
  \bibinfo{author}{\bibfnamefont{C.~A.} \bibnamefont{Baker}},
  \bibinfo{author}{\bibfnamefont{G.}~\bibnamefont{Ban}},
  \bibinfo{author}{\bibfnamefont{G.}~\bibnamefont{Bison}},
  \bibinfo{author}{\bibfnamefont{K.}~\bibnamefont{Bodek}},
  \bibinfo{author}{\bibfnamefont{V.}~\bibnamefont{Bondar}},
  \bibinfo{author}{\bibfnamefont{M.}~\bibnamefont{Burghoff}},
  \bibinfo{author}{\bibfnamefont{E.}~\bibnamefont{Chanel}},
  \bibnamefont{et~al.}, \bibinfo{journal}{Phys. Rev. Lett.}
  \textbf{\bibinfo{volume}{124}}, \bibinfo{pages}{081803}
  (\bibinfo{year}{2020}).

\bibitem[{\citenamefont{Graner et~al.}(2016)\citenamefont{Graner, Chen,
  Lindahl, and Heckel}}]{Graner2016}
\bibinfo{author}{\bibfnamefont{B.}~\bibnamefont{Graner}},
  \bibinfo{author}{\bibfnamefont{Y.}~\bibnamefont{Chen}},
  \bibinfo{author}{\bibfnamefont{E.~G.} \bibnamefont{Lindahl}},
  \bibnamefont{and} \bibinfo{author}{\bibfnamefont{B.~R.}
  \bibnamefont{Heckel}}, \bibinfo{journal}{Phys. Rev. Lett.}
  \textbf{\bibinfo{volume}{116}}, \bibinfo{pages}{161601}
  (\bibinfo{year}{2016}).

\bibitem[{\citenamefont{Sachdeva et~al.}(2019)\citenamefont{Sachdeva, Fan,
  Babcock, Burghoff, Chupp, Degenkolb, Fierlinger, Haude, Kraegeloh, Kilian
  et~al.}}]{Xe1}
\bibinfo{author}{\bibfnamefont{N.}~\bibnamefont{Sachdeva}},
  \bibinfo{author}{\bibfnamefont{I.}~\bibnamefont{Fan}},
  \bibinfo{author}{\bibfnamefont{E.}~\bibnamefont{Babcock}},
  \bibinfo{author}{\bibfnamefont{M.}~\bibnamefont{Burghoff}},
  \bibinfo{author}{\bibfnamefont{T.~E.} \bibnamefont{Chupp}},
  \bibinfo{author}{\bibfnamefont{S.}~\bibnamefont{Degenkolb}},
  \bibinfo{author}{\bibfnamefont{P.}~\bibnamefont{Fierlinger}},
  \bibinfo{author}{\bibfnamefont{S.}~\bibnamefont{Haude}},
  \bibinfo{author}{\bibfnamefont{E.}~\bibnamefont{Kraegeloh}},
  \bibinfo{author}{\bibfnamefont{W.}~\bibnamefont{Kilian}},
  \bibnamefont{et~al.}, \bibinfo{journal}{Phys. Rev. Lett.}
  \textbf{\bibinfo{volume}{123}}, \bibinfo{pages}{143003}
  (\bibinfo{year}{2019}).

\bibitem[{\citenamefont{Allmendinger et~al.}(2019)\citenamefont{Allmendinger,
  Engin, Heil, Karpuk, Krause, Niederl\"ander, Offenh\"ausser, Repetto,
  Schmidt, and Zimmer}}]{Xe2}
\bibinfo{author}{\bibfnamefont{F.}~\bibnamefont{Allmendinger}},
  \bibinfo{author}{\bibfnamefont{I.}~\bibnamefont{Engin}},
  \bibinfo{author}{\bibfnamefont{W.}~\bibnamefont{Heil}},
  \bibinfo{author}{\bibfnamefont{S.}~\bibnamefont{Karpuk}},
  \bibinfo{author}{\bibfnamefont{H.-J.} \bibnamefont{Krause}},
  \bibinfo{author}{\bibfnamefont{B.}~\bibnamefont{Niederl\"ander}},
  \bibinfo{author}{\bibfnamefont{A.}~\bibnamefont{Offenh\"ausser}},
  \bibinfo{author}{\bibfnamefont{M.}~\bibnamefont{Repetto}},
  \bibinfo{author}{\bibfnamefont{U.}~\bibnamefont{Schmidt}}, \bibnamefont{and}
  \bibinfo{author}{\bibfnamefont{S.}~\bibnamefont{Zimmer}},
  \bibinfo{journal}{Phys. Rev. A} \textbf{\bibinfo{volume}{100}},
  \bibinfo{pages}{022505} (\bibinfo{year}{2019}).

\bibitem[{\citenamefont{Sarriguren et~al.}(1996)\citenamefont{Sarriguren,
  Moya~de Guerra, and Nojarov}}]{nuc-res}
\bibinfo{author}{\bibfnamefont{P.}~\bibnamefont{Sarriguren}},
  \bibinfo{author}{\bibfnamefont{E.}~\bibnamefont{Moya~de Guerra}},
  \bibnamefont{and} \bibinfo{author}{\bibfnamefont{R.}~\bibnamefont{Nojarov}},
  \bibinfo{journal}{Phys. Rev. C} \textbf{\bibinfo{volume}{54}},
  \bibinfo{pages}{690} (\bibinfo{year}{1996}).

\bibitem[{\citenamefont{Szmytkowski}(2007)}]{Recurrence}
\bibinfo{author}{\bibfnamefont{R.}~\bibnamefont{Szmytkowski}},
  \bibinfo{journal}{J. Math. Chem.} \textbf{\bibinfo{volume}{42}},
  \bibinfo{pages}{397} (\bibinfo{year}{2007}).

\bibitem[{\citenamefont{Nilsson}(1955)}]{Nilsson}
\bibinfo{author}{\bibfnamefont{S.~G.} \bibnamefont{Nilsson}},
  \bibinfo{journal}{Dan. Mat. Fys. Medd.} \textbf{\bibinfo{volume}{29}},
  \bibinfo{pages}{1} (\bibinfo{year}{1955}).

\bibitem[{\citenamefont{Berestetskii et~al.}(1982)\citenamefont{Berestetskii,
  Lifshitz, and Pitaevskii}}]{Landau4}
\bibinfo{author}{\bibfnamefont{V.~B.} \bibnamefont{Berestetskii}},
  \bibinfo{author}{\bibfnamefont{E.~M.} \bibnamefont{Lifshitz}},
  \bibnamefont{and} \bibinfo{author}{\bibfnamefont{L.~P.}
  \bibnamefont{Pitaevskii}}, \emph{\bibinfo{title}{Quantum electrodynamics}},
  vol.~\bibinfo{volume}{4} (\bibinfo{publisher}{Butterworth-Heinemann},
  \bibinfo{year}{1982}).

\bibitem[{\citenamefont{Greiner}(2000)}]{greiner2000relativistic}
\bibinfo{author}{\bibfnamefont{W.}~\bibnamefont{Greiner}},
  \emph{\bibinfo{title}{Relativistic quantum mechanics: Wave Equations}},
  vol.~\bibinfo{volume}{2} (\bibinfo{publisher}{Springer},
  \bibinfo{year}{2000}).

\bibitem[{\citenamefont{Pachucki et~al.}(1993)\citenamefont{Pachucki,
  Leibfried, and H\"ansch}}]{Pachucki93}
\bibinfo{author}{\bibfnamefont{K.}~\bibnamefont{Pachucki}},
  \bibinfo{author}{\bibfnamefont{D.}~\bibnamefont{Leibfried}},
  \bibnamefont{and} \bibinfo{author}{\bibfnamefont{T.~W.}
  \bibnamefont{H\"ansch}}, \bibinfo{journal}{Phys. Rev. A}
  \textbf{\bibinfo{volume}{48}}, \bibinfo{pages}{R1} (\bibinfo{year}{1993}).

\end{thebibliography}

\begin{widetext}
\begin{center}
\begin{table}[htb]
\begin{tabular}{|c|c|c|c|c|c|c|c|c|c|}
\hline
\multirow{3}{*}{} & \multicolumn{2}{c}{Proton transitions} & 
\multicolumn{2}{|c|}{Neutron transitions} & \multirow{3}{*}{$\langle {\bf l}\cdot {\bf s}\rangle_p$} & \multirow{3}{*}{$\langle {\bf l}\cdot {\bf s}\rangle_n$} & $R_0$ & $\Delta \bar E$ & \multirow{3}{*}{$\delta$} \\ \cline{2-5}
& \multirow{2}{*}{ $|\langle n' |{\bf s}|0'\rangle_p |^2$ } & $\Delta E_{n'}$ & \multirow{2}{*}{$|\langle n' |{\bf s}|0\rangle_n' |^2$} & $\Delta E_{n'}$ & & & (fm) & (MeV) & \\
& & (MeV) &  & (MeV) & & & & &\\ \hline
\multirow{7}{*}{$^{138}$Ba} & 18/25 & 2.7 & 170/121 & 5.3 & \multirow{7}{*}{7} & \multirow{7}{*}{15} & \multirow{7}{*}{6.20} & \multirow{7}{*}{14.8} & \multirow{7}{*}{0.09}\\
                            						  & 2/25    & 4.1 & 200/121 & 5.4 & & & & &\\
                            						  & 28/81   & 4.3 & 30/121  & 5.5 & & & & &\\
                           							  & 56/81   & 4.4 & 136/121 & 5.9 & & & & &\\
                            						  & 16/81   & 4.5 & 56/121  & 6.0 & & & & &\\
                            						  & 8/9     & 4.6 & 60/121  & 6.2 & & & & &\\
                            						  & 8/81    & 5.2 & 8/121   & 6.5 & & & & &\\
\hline
\multirow{9}{*}{$^{206}$Pb} & 10/11   & 4.5 & 72/169  & 6.1 & \multirow{9}{*}{15} & \multirow{9}{*}{22} & \multirow{9}{*}{7.09} & \multirow{9}{*}{13.3} & \multirow{9}{*}{0.03} \\
                                                      & 162/121 & 4.6 & 462/169 & 6.2 & & & & &\\
                                                      & 98/121  & 4.7 & 318/169 & 6.3 & & & & &\\
                            						  & 250/121 & 4.8 & 132/169 & 6.4 & & & & &\\
                            						  & 32/121  & 5.0 & 100/169 & 6.5 & & & & &\\
                            						  & 8/121   & 5.1 & 6/169   & 6.7 & & & & &\\
                           							  &         &     & 2/169   & 6.9 & & & & &\\
                            						  &         &     & 2/3     & 1.4 & & & & &\\
                           							  &         &     & 10/9    & 2.0 & & & & &\\
\hline
\multirow{7}{*}{$^{208}$Pb} & 10/11   & 4.5 & 72/169  & 6.1 & \multirow{7}{*}{15} & \multirow{7}{*}{21} & \multirow{7}{*}{7.11} & \multirow{7}{*}{13.3} & \multirow{7}{*}{0.05} \\
                            						  & 162/121 & 4.6 & 462/169 & 6.2 & & & & &\\
                           							  & 98/121  & 4.7 & 318/169 & 6.3 & & & & &\\
                            						  & 250/121 & 4.8 & 132/169 & 6.4 & & & & & \\
                            						  & 32/121  & 5.0 & 100/169 & 6.5 & & & & &\\
                           							  & 8/121   & 5.1 & 6/169   & 6.7 & & & & &\\
                            						  &         &     & 2/169   & 6.9 & & & & &\\
\hline
\end{tabular}
\caption{Nuclear radii $R_0$, deformation parameters $\delta$, nucleon spin-orbit expectation values $\langle {\bf l}\cdot {\bf s}\rangle_{p,n}$, E1 giant resonance energies $\Delta\bar E$ and matrix elements $|\langle n' |{\bf s}|0'\rangle_{p,n} |^2$ and energies $\Delta E_{n'}$ of M1 spin-flip transitions in some spherical nuclei of interest.}\label{sphericalNuc}
\end{table}
%\end{center}
%\end{widetext}

%\begin{widetext}
%\begin{center}
\begin{table}[htb]
\begin{tabular}{|c|c|c|c|c|c|c|c|c|c|}
\hline
\multirow{3}{*}{} & \multicolumn{2}{c}{Proton transitions} & 
\multicolumn{2}{|c|}{Neutron transitions} & \multirow{3}{*}{$\langle {\bf l}\cdot {\bf s}\rangle_p$} & \multirow{3}{*}{$\langle {\bf l}\cdot {\bf s}\rangle_n$} & $R_0$ & $\Delta \bar E$ & \multirow{3}{*}{$\delta$} \\ \cline{2-5}
& \multirow{2}{*}{Transition} & $\Delta E_{n'}$ & \multirow{2}{*}{Transition} & $\Delta E_{n'}$ & & & (fm) & (MeV) & \\
& & (MeV) &  & (MeV) & & & & &\\ \hline
\multirow{5}{*}{$^{172}$Yb} & $\ket{523\frac72} \to  \ket{\frac52} $ & 4.5 & $ \ket{651\frac{3}2} \to  \ket{\frac12}$ & 3.9 & \multirow{5}{*}{10} & \multirow{5}{*}{15} & \multirow{5}{*}{6.67} & \multirow{5}{*}{14.0} & \multirow{5}{*}{0.31} \\
                                                      & $\ket{532\frac52} \to  \ket{\frac32} $ & 4.0 & $ \ket{642\frac52} \to  \ket{\frac32}$                  & 4.5 & & & & & \\
                                                      & $\ket{541\frac32} \to  \ket{\frac12} $ & 4.5 & $ \ket{633\frac72} \to  \ket{\frac52}$                  & 5.0  & & & & &\\
                                                      & $\ket{404\frac92} \to  \ket{\frac72} $ & 4.1 & $ \ket{505\frac{11}2} \to  \ket{\frac92}$               & 5.1 & & & & & \\
                                                      &                                        &     & $ \ket{514\frac{9}2} \to  \ket{\frac{7}2}$              & 4.6 & & & & & \\
\hline
\multirow{6}{*}{$^{174}$Yb} & $\ket{523\frac72} \to  \ket{\frac52} $ & 4.5 & $ \ket{651\frac{3}2} \to  \ket{\frac12}$ & 3.9 & \multirow{6}{*}{10} & \multirow{6}{*}{17} & \multirow{6}{*}{6.70} & \multirow{6}{*}{14.0} & \multirow{6}{*}{0.31} \\
                                                      & $\ket{532\frac52} \to  \ket{\frac32} $ & 4.0 & $ \ket{642\frac52} \to  \ket{\frac32}$                  & 4.5 & & & & & \\
                                                      & $\ket{541\frac32} \to  \ket{\frac12} $ & 4.5 & $ \ket{633\frac72} \to  \ket{\frac52}$                  & 5.0 & & & & & \\
                                                      & $\ket{404\frac92} \to  \ket{\frac72} $ & 4.1 & $ \ket{505\frac{11}2} \to  \ket{\frac92}$               & 5.1 & & & & & \\
                                                      &                                        &     & $ \ket{514\frac{9}2} \to  \ket{\frac{7}2}$              & 4.6 & & & & & \\
                                                      &                                        &     & $ \ket{512\frac{5}2} \to  \ket{\frac{3}2}$              & 2.4 & & & & & \\
\hline
\multirow{5}{*}{$^{176}$Yb} & $\ket{523\frac72} \to  \ket{\frac52} $ & 4.5 &
    $ \ket{651\frac{3}2} \to  \ket{\frac12}$ & 4.2 & \multirow{5}{*}{10} & \multirow{5}{*}{13} & \multirow{5}{*}{6.72} & \multirow{5}{*}{13.9} & \multirow{5}{*}{0.29} \\
                                                      & $\ket{532\frac52} \to  \ket{\frac32} $ & 4.1 & $ \ket{642\frac52} \to  \ket{\frac32}$                  & 4.5 & & & & & \\
													  & $\ket{541\frac32} \to  \ket{\frac12} $ & 4.5 & $ \ket{633\frac72} \to  \ket{\frac52}$                  & 5.0 & & & & & \\
													  & $\ket{404\frac92} \to  \ket{\frac72} $ & 4.0 & $ \ket{505\frac{11}2} \to  \ket{\frac92}$               & 5.2 & & & & & \\
													  &                                        &     & $ \ket{512\frac{5}2} \to  \ket{\frac{3}2}$              & 2.4 & & & & & \\
\hline
\multirow{5}{*}{$^{178}$Hf} & $\ket{523\frac72} \to  \ket{\frac52} $ & 4.4 & $ \ket{505\frac{11}2} \to  \ket{\frac92}$ & 4.2 & \multirow{5}{*}{9} & \multirow{5}{*}{13} & \multirow{5}{*}{6.75} & \multirow{5}{*}{13.8} & \multirow{5}{*}{0.26} \\
                                                      & $\ket{532\frac52} \to  \ket{\frac32} $ & 4.1 & $ \ket{512\frac52} \to  \ket{\frac32}$                  & 2.4 & & & & & \\
													  & $\ket{541\frac32} \to  \ket{\frac12} $ & 4.1 & $ \ket{633\frac72} \to  \ket{\frac52}$                  & 5.0 & & & & & \\
													  & $\ket{402\frac52} \to  \ket{\frac32} $ & 1.9 & $ \ket{642\frac52} \to  \ket{\frac32}$                  & 4.6 & & & & & \\
												      & $\ket{411\frac32}\to\ket{\frac12}$     & 1.4 & $ \ket{631\frac{3}2} \to  \ket{\frac12}$                & 7.8 & & & & & \\
\hline
\multirow{6}{*}{$^{180}$Hf} & $\ket{523\frac72} \to  \ket{\frac52} $ & 4.4 & $ \ket{505\frac{11}2} \to  \ket{\frac92}$ & 4.2 & \multirow{6}{*}{9} & \multirow{6}{*}{17} & \multirow{6}{*}{6.80} & \multirow{6}{*}{13.8} & \multirow{6}{*}{0.25} \\
													  & $\ket{532\frac52} \to  \ket{\frac32} $ & 4.1 & $ \ket{512\frac52} \to  \ket{\frac32}$                  & 2.4 & & & & & \\
													  & $\ket{541\frac32} \to  \ket{\frac12} $ & 4.1 & $ \ket{624\frac92} \to  \ket{\frac72}$                  & 5.3 & & & & & \\
													  & $\ket{402\frac52} \to  \ket{\frac32} $ & 1.9 & $ \ket{633\frac72} \to  \ket{\frac52}$ 				  & 5.0 & & & & & \\
													  & $\ket{411\frac32}\to\ket{\frac12}$     & 1.4 & $ \ket{642\frac{5}2} \to  \ket{\frac32}$                & 4.6 & & & & & \\
 													  &                                        &     & $ \ket{631\frac{3}2} \to  \ket{\frac{1}2}$              & 7.8 & & & & & \\
\hline
\multirow{4}{*}{$^{226}$Ra} & $\ket{523\frac72} \to  \ket{\frac52} $ & 4.3 & $ \ket{624\frac{9}2} \to  \ket{\frac72}$ & 5.0 & \multirow{4}{*}{12} & \multirow{4}{*}{16} & \multirow{4}{*}{7.31} & \multirow{4}{*}{13.0} & \multirow{4}{*}{0.20} \\
													  & $\ket{514\frac92} \to  \ket{\frac72} $ & 4.4 & $ \ket{615\frac{11}2} \to  \ket{\frac92}$               & 5.0 & & & & & \\
													  & $\ket{505\frac{11}2} \to\ket{\frac92}$ & 4.4 & $ \ket{606\frac{13}2} \to  \ket{\frac{11}2}$            & 5.6 & & & & & \\
 													  &                                        &     & $ \ket{761\frac{3}2} \to  \ket{\frac12}$ 				  & 4.3 & & & & & \\
\hline
\multirow{6}{*}{$^{232}$Th} & $\ket{651\frac32} \to  \ket{\frac12} $  & 4.5 & $ \ket{752\frac52} \to  \ket{\frac32}$ & 4.1 & \multirow{6}{*}{13} & \multirow{6}{*}{19} & \multirow{6}{*}{7.37} & \multirow{6}{*}{12.9} & \multirow{6}{*}{0.25} \\
													  & $\ket{505\frac{11}2} \to  \ket{\frac92} $ & 4.2 & $ \ket{761\frac32} \to  \ket{\frac12}$                  & 4.0 & & & & & \\
													  & $\ket{514\frac92} \to  \ket{\frac72} $    & 4.0 & $ \ket{631\frac32} \to  \ket{\frac12}$                  & 1.0 & & & & & \\
													  & $\ket{523\frac72} \to  \ket{\frac52} $    & 3.7 & $ \ket{624\frac92} \to  \ket{\frac72}$                  & 5.0 & & & & & \\
													  &                                           &     & $ \ket{615\frac{11}2} \to  \ket{\frac92}$               & 4.8 & & & & & \\
													  &  										  &  	& $ \ket{606\frac{13}2} \to  \ket{\frac{11}2}$            & 5.4 & & & & & \\
\hline
\end{tabular}
\caption{Nuclear radii $R_0$, deformation parameters $\delta$, nucleon spin-orbit expectation values $\langle {\bf l}\cdot {\bf s}\rangle_{p,n}$, E1 giant resonance energies $\Delta\bar E$ and matrix elements $|\langle n' |{\bf s}|0'\rangle_{p,n} |^2$ and energies $\Delta E_{n'}$ of M1 spin-flip transitions in some deformed nuclei of interest.}\label{deformNuc}
\end{table}
%\end{center}
%\end{widetext}

%\begin{widetext}
%\begin{center}
\begin{table}[htb]
\renewcommand{\arraystretch}{1.2}
\begin{tabular}{|c|c|c|c|c|c|c|c|c|c|c|}
\hline
\multirow{2}{*}{} & \multicolumn{3}{|c|}{Spherical} & \multicolumn{7}{|c|}{Deformed}\\
\cline{2-11}
                  & ${}^{138}{\rm Ba}$ & ${}^{206}{\rm Pb}$ & ${}^{208}{\rm Pb}$ & ${}^{172}{\rm Yb}$ & ${}^{174}{\rm Yb}$ & ${}^{176}{\rm Yb}$ & ${}^{178}{\rm Hf}$ & ${}^{180}{\rm Hf}$ & ${}^{226}{\rm Ra}$ & ${}^{232}{\rm Th}$ \\
\hline
$M_p/a_B$             & 11.1 & 94.1 & 94.0  & 69.3 & 69.3 & 69.2 & 121  & 121  & 156  & 244 \\
\hline
$M_n/a_B$             & 16.5 & 143  & 95.7  & 83.4 & 106  & 88.7 & 96.8 & 114  & 196  & 385 \\
\hline
$\bar{M}/a_B$        & 1.83 & 10.7 & 10.7  & 4.83 & 4.83 & 4.74 & 5.53 & 5.54 & 16.0 & 18.3 \\
\hline
\end{tabular}
\caption{\label{Electronic}
Numerical values for the electronic matrix elements $M_p$, $M_n$ and $\bar M$ for several atoms of interest.}
\end{table}

\end{center}
\end{widetext}

\end{document}